\documentclass[iop,appendixfloats]{emulateapj}

\newcommand{\figurepath}{.}

\newcommand{\bit}[1]{ \textbf{\emph{#1}} }

\newcommand{\ri}{r_{\rm 0}}
\newcommand{\Ri}{R_{\rm 0}}

\newcommand{\Mr}{M_{R}}

\newcommand{\Macc}{\dot{M}_{\rm acc}}

\newcommand{\etatens}{\overline{\overline{\eta}}}
\newcommand{\alphatens}{ {\overline{\overline{\alpha}}}_{\rm dyn} }

\newcommand{\Grav}{\Phi_{\rm g}}

\newcommand{\Cs}{C_{\rm s}}

\newcommand{\Vtot}{\bit{V}}

\newcommand{\Vr}{V_{R}}
\newcommand{\Vth}{V_{\rm \theta}}
\newcommand{\Vphi}{V_{\rm \phi}}

\newcommand{\Btot}{\bit{B}}
\newcommand{\Bp}{B_{\rm P}}
\newcommand{\Br}{B_{\rm R}}
\newcommand{\Bth}{B_{ \theta}}
\newcommand{\Bphi}{B_{ \phi}}

\newcommand{\Jtot}{\bit{J}}

\newcommand{\ass}{{\alpha}_{\rm ssm}}
\newcommand{\am}{{\alpha}_{\rm m}}
\newcommand{\adyn}{{\alpha}_{\rm dyn}}
\newcommand{\azero}{{\alpha}_0}

\newcommand{\Fa}{ F_{\rm \alpha} }

\newcommand{\MUseed}{ \mu_{\rm seed} }
\newcommand{\MUzero}{ \mu_{\rm 0} }
\newcommand{\MUact}{ \mu_{\rm act} }
\newcommand{\MUdisk}{ \mu_{\rm D} }
\newcommand{\MUinit}{\mu_{\rm init}}

\newcommand{\ALdisk}{ \alpha_{\rm D} }

\newbox\grsign \setbox\grsign=\hbox{$>$}
\newdimen\grdimen \grdimen=\ht\grsign
\newbox\laxbox \newbox\gaxbox
\setbox\gaxbox=\hbox{\raise.5ex\hbox{$>$}\llap
     {\lower.5ex\hbox{$\sim$}}}\ht1=\grdimen\dp1=0pt
\setbox\laxbox=\hbox{\raise.5ex\hbox{$<$}\llap
     {\lower.5ex\hbox{$\sim$}}}\ht2=\grdimen\dp2=0pt

\shorttitle{Accretion-ejection dynamo}
\shortauthors{Stepanovs et al.}

\usepackage{ctable} 
\usepackage{epstopdf}
\usepackage{pbox}

\begin{document}


\title{Modelling MHD accretion-ejection - 
         episodic ejections of jets triggered by a mean-field disk dynamo}

\author{Deniss Stepanovs, Christian Fendt, Somayeh Sheikhnezami}
\affil{Max Planck Institute for Astronomy, K\"onigstuhl 17, D-69117 Heidelberg, Germany}
\email{deniss@stepanovs.org, fendt@mpia.de}

\begin{abstract}
We present MHD simulations exploring the launching, acceleration and collimation of jets
and disk winds. 
The evolution of the disk structure is consistently taken into account.
Extending our earlier studies, we now consider the self-generation of the magnetic 
field by an $\alpha^2\Omega$ mean-field dynamo.
The disk magnetization remains on a rather low level,
that helps to evolve the simulations for $T > 10,000$ dynamical time steps
on a domain extending 1500 inner disk radii.
We find a magnetic field of the inner disk similar to the commonly found open 
field structure, favoring magneto-centrifugal launching.
The outer disk field is highly inclined and predominantly radial. 
Here, differential rotation induces a strong toroidal component that plays a key role in outflow launching.
These outflows from the outer disk are slower, denser, and less collimated.
If the dynamo action is not quenched, magnetic flux is continuously generated,
diffuses outward the disk, and fills the entire disk.
We have invented a toy model triggering a time-dependent mean-field dynamo.
The duty cycles of this dynamo lead to episodic ejections on similar timescales.
When the dynamo is suppressed as the magnetization falls below a critical value, the generation 
of the outflows and also accretion is inhibited.
The general result is that we can steer episodic ejection and {\em large-scale jet knots}
by a {\em disk-intrinsic dynamo} that is time-dependent and regenerates the jet-launching 
magnetic field.
\end{abstract}

\keywords{accretion, accretion disks --
   MHD -- 
   ISM: jets and outflows --
   stars: mass loss --
   stars: pre-main sequence 
   galaxies: jets
 }
\section{Introduction}
%
%
Astrophysical jets as highly collimated beams of high- velocity material and outflows of 
a small degree of collimation and lower speed are a ubiquitous phenomenon in astrophysical
sources of rather different ranges in energy output and a physical extension.
Previous calculations have shown that jets and winds could be produced by the interplay of 
large scale magnetic fields with the accretion disk 
\citep{1982MNRAS.199..883B, 1992ApJ...394..117P, 1997A&A...319..340F}.

Two main mechanisms compete in the acceleration of the material that is lifted from the disk into the outflow.
In addition to the classical magneto-centrifugal acceleration mechanism proposed by \citet{1982MNRAS.199..883B} that is usually compared to a sling-shot mechanism of material along poloidally dominated magnetic field lines, 
acceleration may also be driven by a pressure gradient of the toroidal magnetic field, 
comparable to a mechanical spring mechanism.
This mechanism has been studied extensively 
both analytically \citep{1994MNRAS.267..146L, 1995MNRAS.275..244L, 1996MNRAS.279..389L} 
and numerically \citep{1995ApJ...439L..39U}.
If the toroidal magnetic field is generated contineously, a inflation of the poloidal field structure results and the material enclosed by the poloidal magnetic loops is accelerated
in vertical direction. 
This kind of jet structure is known as a {\em tower jet}, or 
Poynting-dominated jet.

The jet launching and collimation problem\footnote{In this paper we apply the following notation.
With jet {\em launching} we denote the process that lifts accreting material out of the disk and couples it to a disk wind - the accretion-ejection structure. 
With jet {\em formation} we denote the acceleration and collimation of that slow disk wind from the disk surface to a high velocity of super-escape speed, super-Alf\'enic speed, and possibly super-magnetosonic speed}
is usually being addressed numerically applying 
a large scale initial poloidal magnetic field.
This holds in particular for simulations considering the acceleration and collimation 
process only and assuming the underlying disk as a boundary condition
\citep{1995ApJ...439L..39U, 1997ApJ...482..712O, 1999ApJ...526..631K, 2002A&A...395.1045F,
2006ApJ...651..272F, 2006MNRAS.365.1131P, 2009ApJ...692..346F, 2009ApJ...702..567V, 
2010ApJ...709.1100P}.

Also simulations treating the launching mechanism, i.e. simulations of the accretion-ejection
structure that include the time evolution of the disk dynamics, so far have assumed a global 
large-scale magnetic field as initial condition
\citep{1985PASJ...37...31S,2002ApJ...581..988C, 2004ApJ...601...90C, 2006A&A...460....1M, 
2007A&A...469..811Z, 2009MNRAS.400..820T, 2010A&A...512A..82M, 2012ApJ...757...65S,
2013ApJ...774...12F}.

These studies have provided a deep insight in the launching mechanism, i.e. the connection 
between the outflow and the underlying disk.
It is clear today that the magnetic field plays {\em the} crucial role in lifting
the matter out of the disk and accelerating it to high velocity.
By knowing the disk magnetization one can refer many details of the launched outflow, namely 
its energetics and the ejection efficiency
(see \citealt{2007A&A...469..811Z, 2012ApJ...757...65S, paper-1}, hereafter Paper I; \citealt{paper-2}).

It is still an open issue what the exact structure and the strength of the magnetic field in the disk is, and where its origin is.
Besides a central stellar magnetic field or advection of magnetic field from the
ambient medium, a turbulent dynamo can be a major source of the disk magnetic field
\citep{1981MNRAS.195..881P, 1981MNRAS.195..897P, 1995ApJ...446..741B}.
In order to study the disk dynamo in the context of outflow launching, only a few numerical experiments
were performed in which the magnetic field was generated {\it ab initio}
\citep{2001A&A...370..635B, 2003A&A...398..825V, 2004A&A...420...17V}.
These authors were first to show how accretion disks start producing the outflow if the magnetic
field is amplified by the dynamo to about its equipartition value.

A further motivation for considering a disk dynamo for jet-launching is seemingly the 
time-dependent ejection of the jet material.
For protostellar jets the typical timescales for ejection derived from the observed knot 
separation and jet velocity are in the range of 10-100 years.
The typical timescale of the jet-launching area is, however, about 10-20 days, that is
the Keplerian period close to the inner disk radius.
A time-variable dynamo may be responsible for changing the jet-launching conditions on
longer timescales. 
We may here refer to the dynamo cycle of the Solar magnetic field that is longer than the
Sun's rotation period\footnote{By coincidence, the difference in the respective time 
scales - a magnetic cycle of 22 years and 
a rotational period of 35 days - is comparable to the protostellar jets.}.

Our main concern in this paper is structure and time evolution of the dynamo-generated 
magnetic field,
the launching of outflows by such a disk self-generated magnetic field,
and the interrelation between the dynamo and the episodic ejection of jets, 
possibly leading to the co-called jet knots.
Such a study has not yet been presented in the literature.

Disk dynamos were discussed in the literature concerning episodic accretion and ejection events 
in dwarf novae, and also as a possible physical process to generate MHD instabilities and 
turbulence, allowing for angular momentum transfer and accretion.
\citet{1996ApJ...457..332A} discussed a disk dynamo mechanism in accretion disks as a cause for
dwarf novae eruption, similar to what could probably happen in the jet-launching disks.
\citet{1992MNRAS.259..604T} discuss a disk dynamo action in order to physically produce the magnetic 
disk viscosity. 
However, \citet{1998ApJ...492L..75G} showed that for low Reynolds numbers the MHD disk turbulence 
and angular momentum transport dies out, possibly leading to episodic accretion in dwarf novae. 

\citet{1995MNRAS.276.1179R} discuss a model of a disk dynamo driven by magnetic buoyancy which does
not directly involve a disk turbulence.
This model was re-visited by \citet{2008A&A...490..501J} finding that accretion could in fact 
be established by a Parker instability-driven dynamo.
According to \citet{2008A&A...490..501J} accretion could be based on the interaction of 
Parker and magneto-rotational instabilities. 
In this scenario the vertical component of the magnetic field is generated by Parker 
instability (PI) and serves as a source for magnetorotational instability (MRI). 

Applying a mean-field $\alpha^2\Omega$ dynamo \citep{1980opp..bookR....K} 
we present the step by step evolution of the magnetic field.
In our approach turbulence is being addressed in the mean field approach, and is not
self-consistently generated (e.g. by the MRI).

We also study episodic jet-launching scenarios by means of a simple toy model in which we 
artificially switch on/off the dynamo action.
We discuss whether similar processes in which the dynamo does change its strength may lead 
to episodic jet ejection and the jet knots.
We also discuss in detail how the magnetic field can be regenerated by re-establishing the 
dynamo action.

Our paper is organized as follows. 
In Section~2 we briefly describe our numerical setup. For a more complete discussion we
refer to Paper I.
In particular, we discuss the implementation of the mean-field dynamo equations and the 
model approach for the magnetic diffusivity and the dynamo-$\alpha$.
In Section~3 we present our reference dynamo simulation where the jet-launching magnetic field structure is
dynamo-generated from a weak seed field. 
We discuss the difference between dynamo and non-dynamo simulations.
In Section~4 we present simulations during which the disk dynamo is switched on and off
repeatedly, leading to episodic ejection of the disk material into the collimated outflow.
We summarize our paper in Section~5.

\section{Model Approach}\label{sec:model}
For our numerical simulations, we apply the MHD code 
PLUTO\footnote{Version 4.0, released 2013} \citep{2007ApJS..170..228M},
solving the time-dependent, resistive MHD equations 
on a spherical grid. 
Our simulations have been performed in 2D axisymmetry, applying spherical coordinates $(R,\theta)$.
We refer to $(r,z)$ as cylindrical coordinates.

We have specified the equations considered in detail in Paper I.
Here we stress in particular the induction equation that we have modified in the 
code according to the mean-field dynamo formalism \citep{1980opp..bookR....K},  
\begin{equation}\label{eq:induction}
\frac{\partial \Btot }{\partial t} = 
\nabla\times (\Vtot \times \Btot + \alphatens \Btot -  \etatens \Jtot),
\end{equation}
where the tensor $\alphatens$ describes the $\alpha$-effect of 
the mean-field dynamo, and the tensor $\etatens$
the magnetic diffusivity (see below).

As no physical scales are introduced in the equations we solve, the results of simulations are
presented in non-dimensional units.
Lengths are given in units of $R_0$, corresponding to inner disk radius. 
Velocities are given in units of $V_{\rm K,0}$, corresponding to the Keplerian speed at $R_0$. 
Thus times are given in  $T_0 \equiv R_0 / V_{\rm K,0}$ units.
Note, that ${ 2 \pi T_0}$ corresponds to one rotation at the innermost orbit.
Densities are given in units of $\rho_0$, corresponding to $R_0$. 
Pressure is measured in $P_{0} =  \epsilon^2 \rho_{0} V_{0}^2$, where $\epsilon$ is the ratio of the initial isothermal sound speed to Keplerian speed taken at the disk midplane. All our simulations were performed with $\epsilon=0.1$.

We normalize all variables, namely $R, \rho, V, B$, to their values at the inner disk 
radius $\Ri$.
We thus may apply our scale-free simulations to a variety of jet sources.
For the typical astrophysical scaling of the code units we refer to Table~1 of Paper I.

We apply a numerical grid with equidistant spacing in $\theta$-direction, 
but stretched cell sizes in radial direction, considering $\Delta R = R \Delta\theta$.
Our computational domain of a size $R=[1, 1500 R_0],\theta=[0,\pi/2]$ is 
discretized with $(N_R \times N_\theta)$ grid cells. 
We use a general resolution of $N_\theta = 128$. 
In order to cover a factor 1500 in radius, we apply $N_R = 600$. 
This gives a resolution of 16 cells per disk height ($2 \epsilon$) in the general case. 
We have also performed a resolution study with 1.5 times higher (lower) resolution, thus using
 $900 \times 192$ ($450 \times 64$) cells for the domain, or 26 (11) cells per disk height.

\subsection{Initial Conditions}
%
%
As a measure for the magnetic field strength, we use the magnetization defined as 
the ratio between magnetic and thermal pressure,
\begin{equation}
\mu = \frac{B^2}{2P}
\end{equation}
We have used different prescriptions for the magnetization, however, in all cases
the local magnetic field pressure $B^2/2$ is related to the gas pressure $P$ at 
the midplane.

All dynamo simulations we perform start from a very weak initial magnetization 
$\MUinit = 10^{-5}$. 
Therefore the initial structure of the accretion disk can be obtained as the solution to 
the steady-state force equilibrium equation, neglecting the contribution by the 
Lorentz force,
\begin{equation}
\nabla P  +\rho \nabla \Grav  
- \frac{1}{R} \rho \Vphi^2 
 ({\bit e}_R \sin\theta + {\bit e}_\theta \cos\theta) = 0.
\end{equation}
Assuming a self-similar disk structure this equation can be solved analytically.

All our simulations are initialized with a purely radial magnetic field, confined 
within the disk and defined via the vector potential $\vec B = \nabla \times A \vec{e_\phi}$, 
and
\begin{equation}
A = B_{\rm p,0} r^{-1} e^{-8{z/H}^2}.
\end{equation}
The parameter $B_{\rm p,0} = \epsilon \sqrt{2 \mu_{\rm init} } $ 
denotes the strength of the initial magnetic field,
while $\epsilon = 0.1$ is ratio of isothermal sound\footnote{Note however, that we use adiabatic equation of state} 
to Keplerian speed. 
Although this magnetic field distribution may be considered as somewhat artificial,
we found that it provides a smooth evolution during the initial phase. 
We have also performed simulations starting from a purely toroidal magnetic field as the initial 
condition, leading to very similar results.

In contrast, purely vertical magnetic field would generate strong currents at the disk surface 
region because of the strong initial shear between the rotating disk and the non-rotating corona. 
This would greatly impact the initial evolution of the accretion-ejection structure.
As long as the {\em initial} magnetization $\MUinit$ is low, it does not play 
a substantial role for the initial disk evolution.
This is the result of the exponential evolution of the magnetic field amplification by the 
dynamo.

\subsection{Boundary Conditions}

\begin{table*}
\caption{Inner and Outer Boundary Conditions.}
\begin{center}
    \begin{tabular}{lllllllll}
    ~         & $\rho$       & P    & $\Vr$    & $\Vth$ & $\Vphi$ & $\Br$ &$\Bth$ & $\Bphi$ \\     
    \noalign{\smallskip}    \hline \hline    \noalign{\smallskip}
    Inner disk  &$\sim r^{-3/2}$&$\sim r^{-5/2}$&$\sim r^{-1/2}, \leq 0$&0&$\sim r^{-1/2}$ &Slope&Slope&$\sim r^{-1}$  \\
    Inner corona  &$\sim r^{-3/2}$&$\sim r^{-5/2}$&$0.2 cos(\varphi)$&$0.2 sin(\varphi)$&$\sim r^{-1/2}$ &0&div B =0&$0$  \\

    Outer disk  &$\sim r^{-3/2}$&$\sim r^{-5/2}$ &Outflow, $\le 0$& Outflow& Outflow&div B =0&Outflow&$\sim r^{-1}$  \\    
            
    Outer corona  &$\sim r^{-3/2}$&$\sim r^{-5/2}$ &Outflow, $\geq 0$& Outflow& Outflow&div B =0&Outflow&$\sim r^{-1}$  \\    
    
    Axis  & Sym & Sym & Sym & Anti-sym & Anti-sym & Sym & Anti-sym & Anti-sym   \\        
    
    Equator& Sym & Sym & Sym & Anti-sym & Sym & Anti-sym & Sym & Anti-sym   \\        

    \noalign{\smallskip}\hline \noalign{\smallskip}

\multicolumn{9}{l}{
{\bf Note.} Outflow is the zero gradient condition and the constant slope conditions are marked by ''slope'' in the table.}

    \end{tabular}
\end{center}
\label{tbl:bc}
\end{table*}

The boundary conditions are adapted from Paper I. 
The only change was made for the coronal region of the inner boundary.
Here, we do not allow any magnetic flux to penetrate the inner coronal region and
not only set $\Bphi = 0$, but also $B_R = 0$.
Since the magnetic field vanishes in that area, we therefore prescribe a purely 
radial profile of the inflow into the corona (in contrast to an inflow aligned to the magnetic field
considered previously), keeping the same inflow velocity $\Vr = 0.2$.
We summarize all boundary conditions in Table~\ref{tbl:bc}

Since the magnetic field is suppressed in the inner coronal region, the shear in the area between 
the coronal region and the disk boundary can develop strong electric currents.
This makes the region between the axis and the jet subject to small-scale perturbations,
especially in the runs with high resolution. 
On the other hand, the jet-launching area of the inner disk always shows a smooth, stable, and 
non-fluctuating evolution.

\subsection{Magnetic Diffusivity}
In Paper I we studied in detail models applying both a {\em standard diffusivity} and 
a so-called {\em strong diffusivity}. 
We have shown numerically that the standard diffusivity model is prone to the accretion 
instability.
In the present paper, studying the dynamo action, we also performed simulations using both models. 
These models qualitatively share many similar features. 
Therefore, we will present simulations applying the strong diffusivity model, however,
commenting on differences between diffusivity model.

The main mediator in the magnetic diffusivity models is the magnetization of 
the underlying disk. 
In case of simulations with a substantially strong initial magnetic field (see Paper I), the 
disk magnetization is set by the magnetic field at the disk midplane. 

Since in the dynamo simulations presented here the initial magnetic field does not intersect with
the midplane\footnote{The initial field is purely radial or toroidal}, and may also remain low 
for quite some time, 
the parametrization of the diffusivity model with the magnetization had to be revised.
We keep the same notation as in Paper I for the strong diffusivity model,
\begin{equation}\label{eq:strongdiff}
\ass = \am \sqrt{2\mu_0} \left(\frac{\MUdisk}{\mu_0}\right)^2,
\end{equation}
where the {\em disk magnetization},
\begin{equation}\label{eq:mudisk}
\MUdisk = \frac{<B_{\rm D}>^2}{2P},
\end{equation}
is defined by means of the {\em average} total magnetic field $<B_{\rm D}>$ for a certain radius
within the disk (up to $H$), normalized  to the midplane pressure.
A non-zero magnetic diffusivity allows for reconnection and diffusion of the
magnetic field across the midplane.
An assumption that the magnetic diffusivity is dependent on the total magnetic field strength 
is consistent with the fact that the MRI is excited by both toroidal and
poloidal magnetic field components \citep{2013EAS....62...95F}.

\subsection{The Dynamo Model}
We apply a standard mean-field $\alpha^2\Omega$ dynamo formalism \citep{1980opp..bookR....K}, 
where $\alpha$ represents the dynamo effect by turbulence, and $\Omega$ stands for 
the differential rotation of the plasma.
According to the mean-field dynamo theory, an extra electromotive force term $\alphatens \Btot$ 
enters the induction equation (Equation~\ref{eq:induction}) and is responsible for the 
generation of the magnetic field.
In general, $\alphatens$ is a tensor, however non-diagonal components are less relevant for 
the dynamo process \citep{1997MNRAS.288L..29B,2001A&A...370..635B}, 
in particular when a moderately strong magnetic field is present \citep{2012SSRv..169..123B}. 
Therefore we neglect all non-diagonal components and set the diagonal values equal to 
one parameter, $\adyn$.
The sign of $\adyn$ as well as its number value in real disks has been widely debated 
\citep{1997LNP...487..109B, 2000A&A...353..813R, 2001A&A...374.1035A}. 
It has been shown that in order to get a dipolar structure of the mean magnetic field 
(as opposed to a quadrupolar structure) a negative sign of alpha should be chosen 
\citep{2007MmSAI..78..374B, 2001A&A...370..635B}.

Following a dimensional analysis, we may scale $\adyn$ as the Keplerian velocity, thus applying
\begin{equation}
\adyn = \ALdisk\,r^{-1/2} \Fa(z),
\end{equation}
where the vertical profile of $\alpha$-effect is defined by
\[ \Fa(z) =   \left\{
\begin{array}{ll}
      \sin( \pi  \frac{z}{H} ) & z\leq H \\
      0 & z > H \\
\end{array} 
\right.
\]
Here, $H$ denotes the disk scale height and is approximated as constant in time. 
The profile $\Fa (z)$ restricts the $\alpha$-effect to the disk area.

It is generally believed that in case of the strong magnetic field the dynamo is quenched \citep{2005PhR...417....1B}.
The main reason is that a strong global magnetic field suppresses the turbulence,
and thus the turblulent dynamo.
The quenching is commonly applied by multiplying the $\adyn$-term by a quenching function,
\begin{equation}\label{eq:quen}
Q = \frac{1}{1 + 2\mu_{\rm x}},
\end{equation}
where $\mu_{\rm x}$, in contrast to the $\MUdisk$ is the {\em local} magnetization.
In order to be consistent with and directly affect the resulting magnetic field,
we parameterize $\mu_{\rm x} = q_{\mu}\MUdisk$. 
By setting different $q_{\mu}$ we can limit the magnetic field growth to a certain value.
Typically, we choose a rather high $q_{\mu}$ in order to quench the dynamo already for low 
magnetizations.

However, there is another possibility of limiting the magnetic field strength.
We find that $\MUzero$ in the strong diffusivity model (Equation~\ref{eq:strongdiff}) 
is in fact a good measure for the resulting actual disk magnetization.
This comes from the functional form of the diffusivity profile - any further growth
of the disk magnetization has a strong feedback of the diffusivity (see Paper I).

Both direct dynamo quenching and indirect limiting the magnetization by applying the strong 
diffusivity model lead to the saturation of the magnetic field.
The difference between this approaches is that in case of leaving the dynamo working in the disk,
the magnetic flux is being continuously generated and 
the disk is being filled with the magnetic field.
If the standard diffusivity model is applied, then the dynamo quenching is the only 
mechanism to stop further magnetic field amplification.
Therefore, since we apply the strong diffusivity model in the simulations we present, 
these simulations are run without dynamo quenching.

The expected dynamo number for accretion disks is given by
\begin{equation}
|{D}| = |{C}_{\alpha} {C}_{\omega}|\lesssim \frac{3}{2} {\ass}^{-2},
\end{equation}
\citep{2003A&A...398..825V} 
where ${C}_{\alpha} = (\azero H / \eta_0)$ and ${C}_{\omega} = (|\Delta\Omega|H^2 / \eta_0)$ 
represent the strength of $\alpha$-effect and shear $d\Omega/dr$, respectively.
Since our main concern is the resulting jet-launching magnetic field, we choose the maximum 
dynamo number in order to generate the magnetic field structure as rapid as possible. 
The maximum dynamo number is provided by $\ALdisk = - 0.1$.
Note, that the dynamo number $D$ is strongly dependent on $\ass$.

\section{A Reference Dynamo Simulation}
In this section we discuss simulations applying the dynamo model and the resulting 
configuration of disk-jet system.

\begin{figure*}
\centering
\includegraphics[width=18cm]{\figurepath/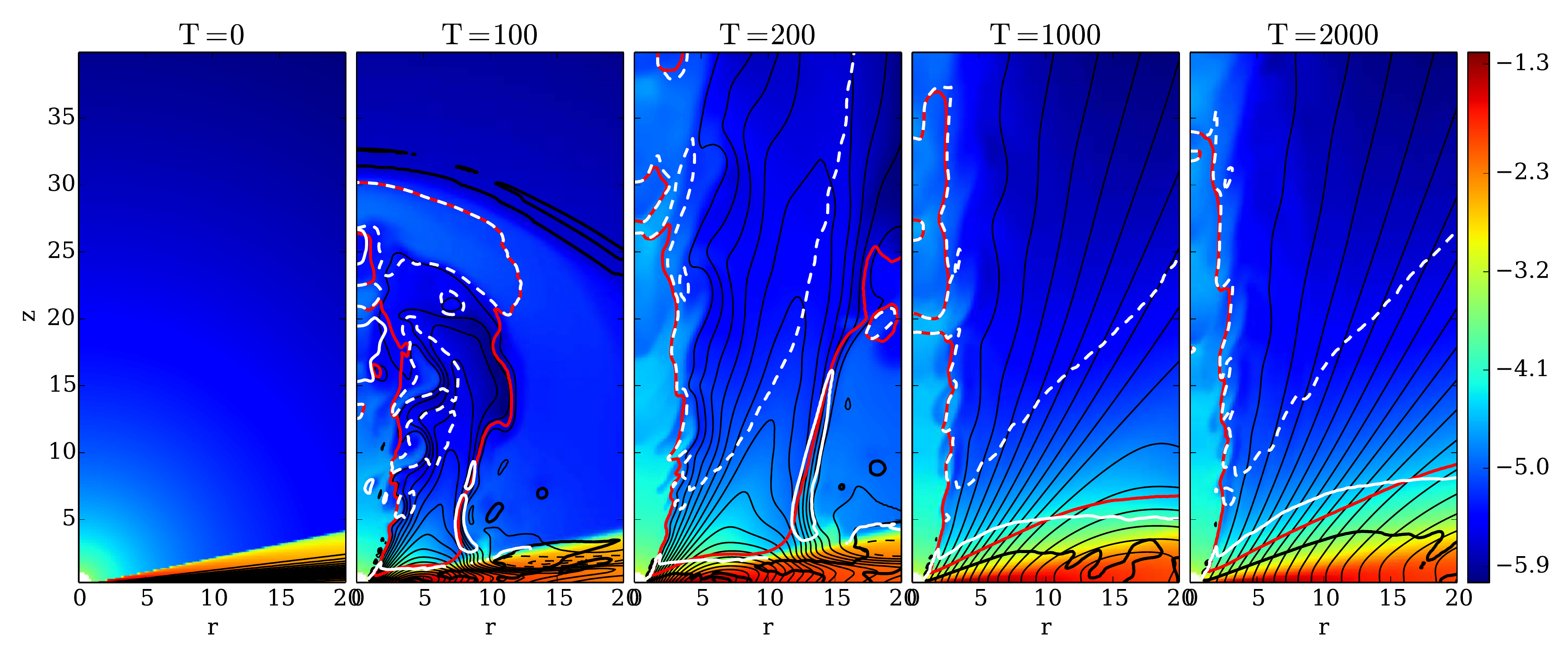}
\caption{Time evolution of the disk-jet structure of the reference dynamo simulation.
Only a small cylindrical part of the whole spherical domain is presented.
Shown is the density (colors, in logarithmic scale);
the poloidal magnetic field lines (thin black lines); 
the disk surface  as defined by $\Vr = 0$ (thick black line);
the sonic (red line), the Alfv\'en (white line), and the fast magnetosonic (white dashed line) surfaces.
}
\label{fig:dyn_tevol}
\end{figure*}

We will refer to our reference dynamo simulation as to the simulation
with the parameters
$\ALdisk = -0.1$, $\am = 1.65$, $\MUzero = 0.01$.
Figure~\ref{fig:dyn_tevol} shows the time evolution of our reference
dynamo simulation, that can be seen as typical for our model setup.
The simulation starts from a weak ($\MUseed = 10^{-5}$), 
purely radial magnetic field, confined within the disk.
Once the simulation is started, the toroidal component of the magnetic field is being 
continuously generated from the radial magnetic field simply by stretching.
For the poloidal magnetic field component, the only generation mechanism is the 
dynamo effect that induces the poloidal component from the toroidal one.

Since the toroidal magnetic field is antisymmetric to equator, the poloidal magnetic field 
loops that are generated by the dynamo first, do {\em not} cross the equator. 
When they evolve in time magnetic reconnection is enforced by the equatorial plane boundary condition
and the magnetic diffusivity in the disk.
As a consequence the magnetic loops (in the upper and lower hemisphere) merge and do traverse the 
equatorial plane.
Since our diffusivity model depends on the average magnetic field in the disk 
there is always a substantial diffusivity present in the disk.

As described by \citet{2008A&A...490..501J}, the toroidal component of the magnetic field 
is continuously amplified
until it reaches the buoyancy limit and starts moving upward, away from the disk.
The upward motion changes the structure of the magnetic field lines from a 
predominantly radial into a vertical direction.
When the magnetic field is sufficiently inclined, magneto-centrifugal launching \citep{1982MNRAS.199..883B}
can strongly accelerated the plasma on these field lines. 
The outflowing gas carries with it the toroidal magnetic field generated in the disk, thus 
setting a limit for the toroidal magnetic field strength in the disk (see Paper I).

All dynamo simulations evolve into three distinct domains in the accretion-ejection 
structure.
Starting from the innermost disk, in the first region the magnetic field has the typical
structure of field lines inclined with respect to the disk surface.
Although magnetic field generation by the dynamo can still take place (if the $\alpha$-effect is not quenched), 
the magnetization in that region has become sufficiently high in order to operate the standard 
magneto-centrifugal jet driving.
The second region is where the poloidal magnetic field is mostly radial. 
Here the velocity shear in the disk creates a strong toroidal component of the magnetic field.
In this area, the outflow is launched mainly by the buoyancy of the toroidal magnetic field.

The third region is the outer disk, where the magnetic field is rather weak, with somewhat
irregular structure. 
Here, the magnetic flux has originated from the same magnetic loops as the innermost field, but 
because of longer distance from the inner disk the magnetic field strength has become much lower.
Note, that the magnetic field in the outer disk has the opposite polarity.

In summary, the overall structure of the magnetic field has a strong gradient
that leads to strong outward diffusion of magnetic flux. 
Due to this diffusion, the whole disk is filled by a non-zero 
net magnetic field ($\Bp|_{\rm midplane}\neq 0$).

\subsection{Dynamo Effect Versus Magnetic Diffusion}
As discussed in detail in Paper I, the evolution of the disk-jet system is mainly set by the two 
opposite processes, the diffusion and advection of the magnetic field.
It is more complicated to reach this balance. 
In case of dynamo simulations a third process contributes to the induction equation, the dynamo,
and it is more complex to reach to an equilibrium situation.
Nevertheless, the main effects of these processes can be disentangled. 
The dynamo term in the induction equation manifests itself by generating loops of the poloidal 
magnetic field.
Because of the magnetic flux conservation along the magnetic loops, 
the magnetization in the inner disk (inner footpoint of the loop) is always higher than in the 
outer disk (outer footpoint of the loop).
Therefore a strong gradient of the magnetic field develops, that evolves primarily following 
the magnetic diffusivity model.

By smoothing out the gradient of the magnetic field, the diffusivity plays a key role 
in the overall evolution of the magnetic field - 
first, it diffuses the magnetic field outward, thus filling the outer disk with the magnetic flux,
second, at the same time the diffusivity destroys some flux within the magnetic loop
by reconnection.

In general, if the dynamo is not sufficiently strong (in case of low dynamo numbers), 
the generated magnetic field will quickly decay (being diffused) and the magnetization
necessary for jet-launching will not be reached.

\subsection{Structure of the Tower Jet}
\begin{figure}
\centering
\includegraphics[width=9cm]{\figurepath/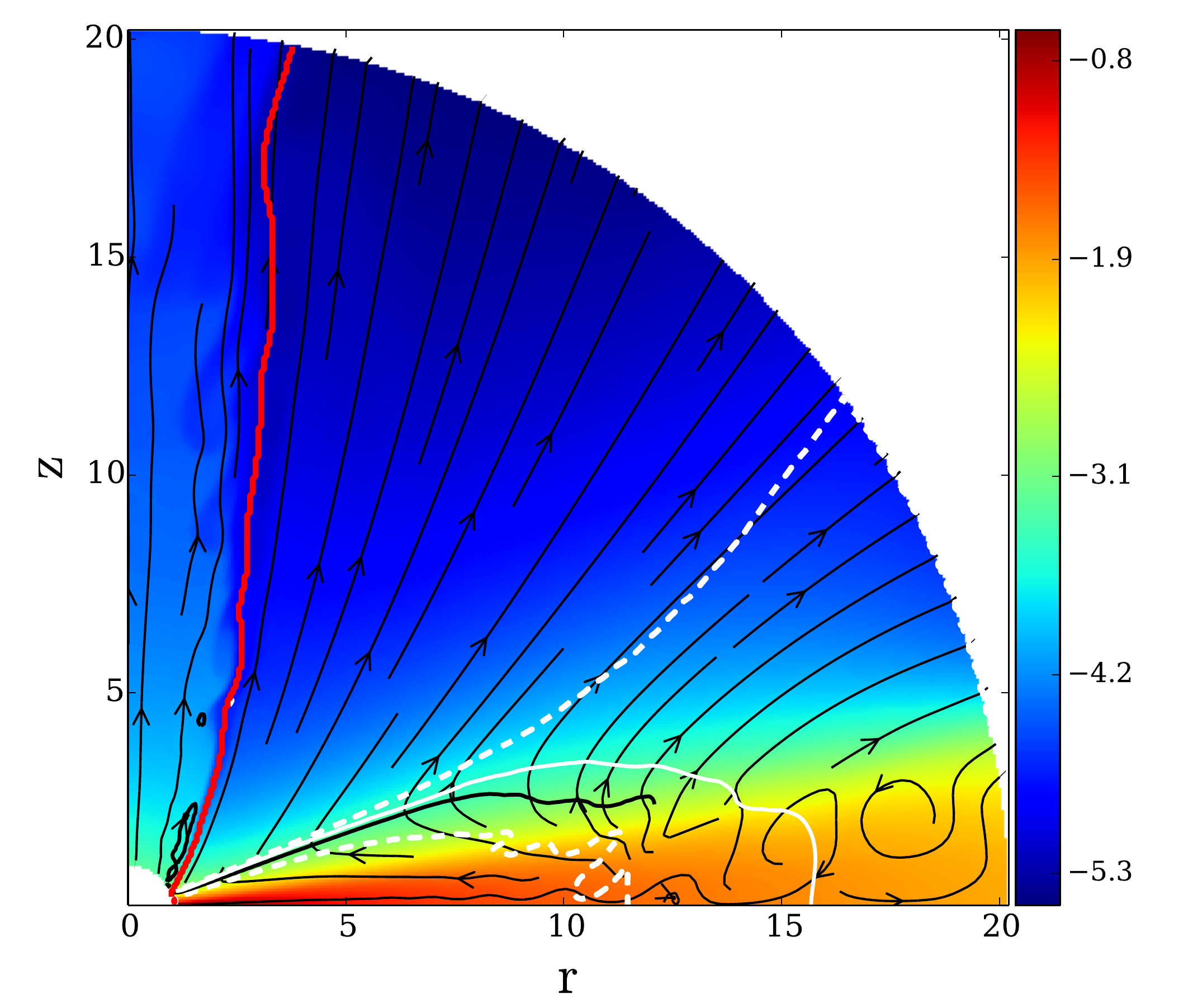}
\caption{Snapshot at $T=1,000$ of physically different regions of the disk-jet structure. 
Shown is the mass density (in logarithmic scale) and
streamlines of the poloidal velocity (black lines with arrows).
The red line marks the magnetic field line that is rooted at the innermost disk area along the 
midplane.
The upper white dashed line separates the area where $V_p || B_p$ from the 
rest of the disk. 
The accretion and ejection areas are separated by a white line indicating $V_r = 0$, and a black line
indicating $F_\phi = 0$, respectively.
The lower white dashed line separates the actual accretion area where $V_r >> V_\theta$ from the rest 
of the disk.}
\label{fig:dyn_struct}
\end{figure}

\begin{figure*}
\centering
\includegraphics[width=18cm]{\figurepath/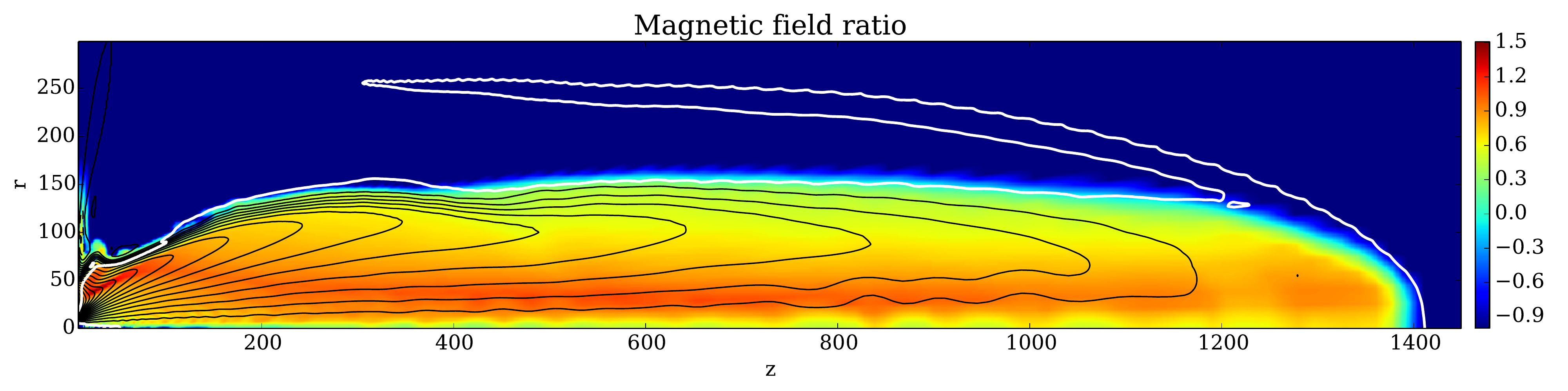}
\includegraphics[width=18cm]{\figurepath/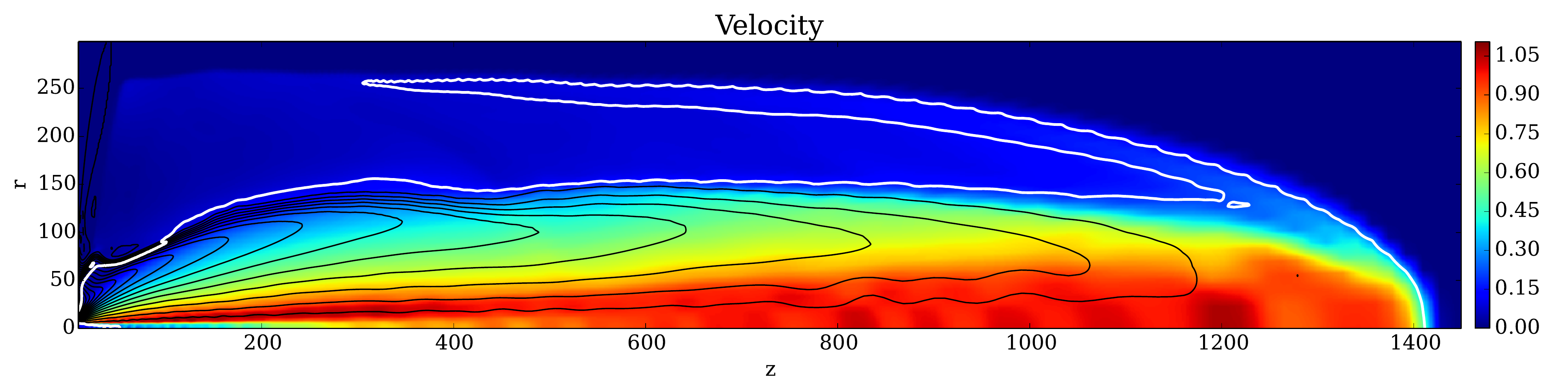}
\caption{Structure of the simulation at $T=1800$.
Shown by the color: logarithm of toroidal to poloidal magnetic field ratio (upper plot) and the jet poloidal speed (lower plot).
The black lines show the magnetic field.
The white line represents the sonic surface.
}
\label{fig:jet_cyl}
\end{figure*}

Figure~\ref{fig:dyn_struct} shows the snapshot of the initial evolution at $T=1,000$.

What can be immediately seen is that the disk structure, namely the structure of the velocity field and the magnetic field (see Figure~\ref{fig:dyn_tevol}) is completely different from the non-dynamo simulations \citep{paper-1,paper-2}.
We find two distinct regions in which the poloidal component of the magnetic field is inclined slightly (for $r<10$) or strongly (for $r>10$) with respect to the disk surface.
While the magnetic field of the inner disk favors a standard magneto-centrifugal launching, in this section, we primarily concentrate mainly on the outer disk.

In this region, the strong toroidal magnetic field is induced by the differential rotation of both the inclined poloidal magnetic field and by the magnetic
loops that are rooted at radially different footpoints along the disk.
The mechanism we observe is similar to the well-known tower jet \citep{1994MNRAS.267..146L, 1995MNRAS.275..244L, 1996MNRAS.279..389L, 1995ApJ...439L..39U}.
The increasing toroidal magnetic field pressure leads to an inflation of the poloidal magnetic loops and the material enclosed by that poloidal loops is accelerated in vertical direction.
This structure -- typical for a tower jet-- is clearly seen in the extended loops in Figure~\ref{fig:jet_cyl}.
In Figure~\ref{fig:jet_cyl}, we further see that the fast jet, the one that is launched from the inner region of the disk, becomes collimated already close to the disk.
On the contrary, for the tower jet - the expanding loop structures - it takes a while to collimate. 

Almost everywhere in the jet region, the toroidal magnetic field dominates the poloidal magnetic field. 
Numerical simulations have shown that such structures, for example, naturally result from the interaction between a stellar dipole magnetic field penetrating the accretion disk 
\citep{1996ApJ...468L..37H, 2000ApJ...541L..21U, 1999A&A...349L..61F, 2000A&A...363..208F, 2004ApJ...600..338K}.
Note, however, that the magnetic loops presented in the disk are generated by the disk dynamo.
The tower jet origins from the magnetic loop structure, and
as the simulation evolves, that magnetic loop structure, and thus the base of the ''tower'', constantly moves outward.

Around the magnetic loop structure ($r \approx 10$), we find that it is the 
the buoyancy force of the toroidal magnetic field that is
the main force responsible for 
the lifting of the disk material into the outflow.
Starting from the disk surface, defined as a surface of zero radial velocity ($\Vr = 0$), the matter is further accelerated 
by the pressure gradient of the toroidal magnetic field.
The latter is, in fact, consistent with the simulations by \citet{1995ApJ...439L..39U}.
The early evolution of the disk-jet system (Figure~\ref{fig:jet_cyl} and Figure~\ref{fig:dyn_tevol} at $T=200$)
clearly show the similarity to the magnetic towers.

We typically find that the launching region, that was defined in Paper I 
as the region where the velocity of the plasma changes from being
perpendicular to the magnetic field to almost parallel to the 
magnetic field, is broader in dynamo simulations than in the non-dynamo 
simulations.
Also, the disk surface, where the radial velocity changes sign by definition,
is located at higher altitudes, although the thermal disk scale height is still 
about constant in time and about its initial value.

\subsection{Outflow Launching: Accretion-ejection}\label{sec:olac}
The magnetic field of the inner disk that is established in the simulations
is similar to the usual structure favoring the magneto-centrifugal launching 
of the outflow.
This type of jet formation 
has been previously found and discussed by many authors 
\citep{1982MNRAS.199..883B, 2004ApJ...601...90C, 2007prpl.conf..277P, 2007A&A...469..811Z, 2010A&A...512A..82M} 
as well in our Paper I.

\begin{figure}
\centering
\includegraphics[width=9cm]{\figurepath/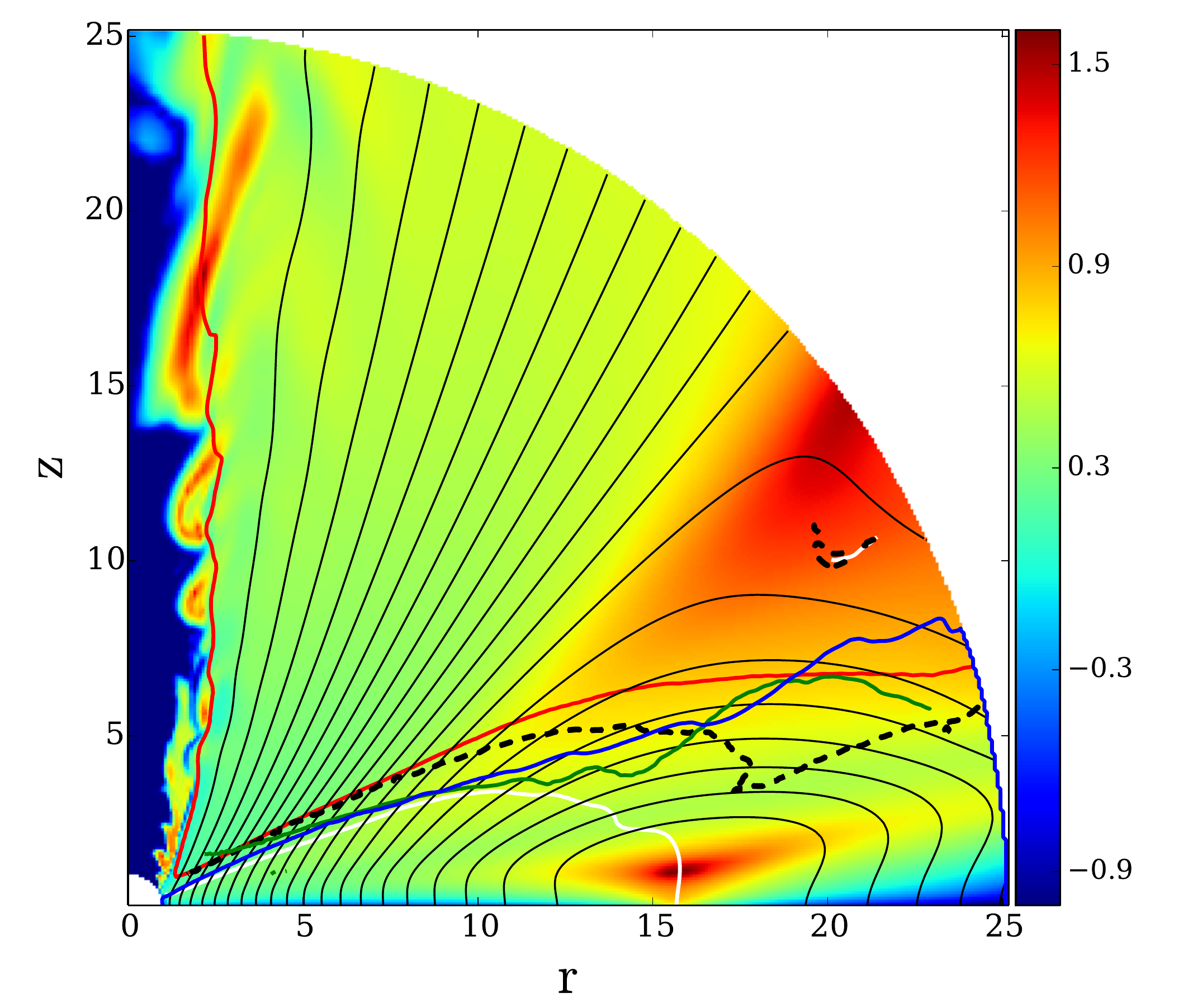}
\caption{Distribution of the magnetic field components at $T=1000$. 
Shown is the ratio of the toroidal to the poloidal magnetic field (colors, {\em logarithmic scale}),
the poloidal magnetic field lines (thin black lines);
the sonic surface (red line); 
and the locations where the Lorentz force components change sign, $F_\phi = 0$ (white line) and 
$F_\theta = 0$ (blue line).
Further denoted are the locations where i) the gas pressure force is equal to the Lorentz force both
projected parallel to the magnetic field (black dashed line), 
and ii) where the gas pressure force is equal to the Lorentz force both projected parallel to the 
velocity field (green line).
}
\label{fig:dyn_btbp}
\end{figure}

Therefore, in this section we concentrate on the (outer) disk region where the
magnetic field has evolved into a structure completely different from the previous simulations.
Namely into the structure with the poloidal magnetic field being predominantly radial, 
and very strong toroidal component.
In this part of the disk, it is the toroidal magnetic field that plays the key role 
in the launching (see Fig.\ref{fig:dyn_btbp}).

As the accretion-ejection process is governed by the magnetic torques, 
these torques need to be discussed in detail.
The white line in Figure~\ref{fig:dyn_btbp} marks the region where the magnetic torque changes sign.
The torque is negative in the inner disk (inside the white line), where the angular momentum extraction
from disk to outflow takes place, and positive in the disk corona, leading to the acceleration of the 
outflow material.
In the region that is dominated by the magnetic loops at radii of $R \simeq 15$ we find that the
torque is purely positive, thus playing a major role in the acceleration of the plasma.
The blue line in Figure~\ref{fig:dyn_btbp} separates two regions, where i) the
magnetic forces accelerate the matter in the direction of the outflow
($F_\theta >0$, above the line), and ii) where the magnetic forces pinch the disk (below the line). 
In the disk area below this line the main force lifting the matter into the outflow
is the thermal pressure.
The lines that mark the area where the pressure force is equal to the Lorentz
force projected parallel to the magnetic field (black dashed line) and parallel to the velocity field (green line)
are also shifted closer toward the disk midplane.

In the area above the loop-like structure the toroidal magnetic field dominates the poloidal field.
In this region acceleration is mainly governed by the toroidal magnetic field pressure gradient.

It is worth to note that such a configuration does not reach a steady state.
This is already indicated by the misalignment between the magnetic field lines and velocity field. 
Furthermore, the blue lines in Figure~\ref{fig:dyn_struct} denote the launching area (see Paper I)
where the velocity field changes from a direction perpendicular to parallel to magnetic field lines.
The longer the simulation evolves, the larger the area that reaches a steady state.
In other words, the non-steady loop structure is moving outward along the disk.

\subsection{Dynamical Profiles of a Dynamo-disk driving jets}
\label{sub:profiles}
\begin{figure}
\centering
\includegraphics[width=8.5cm]{\figurepath/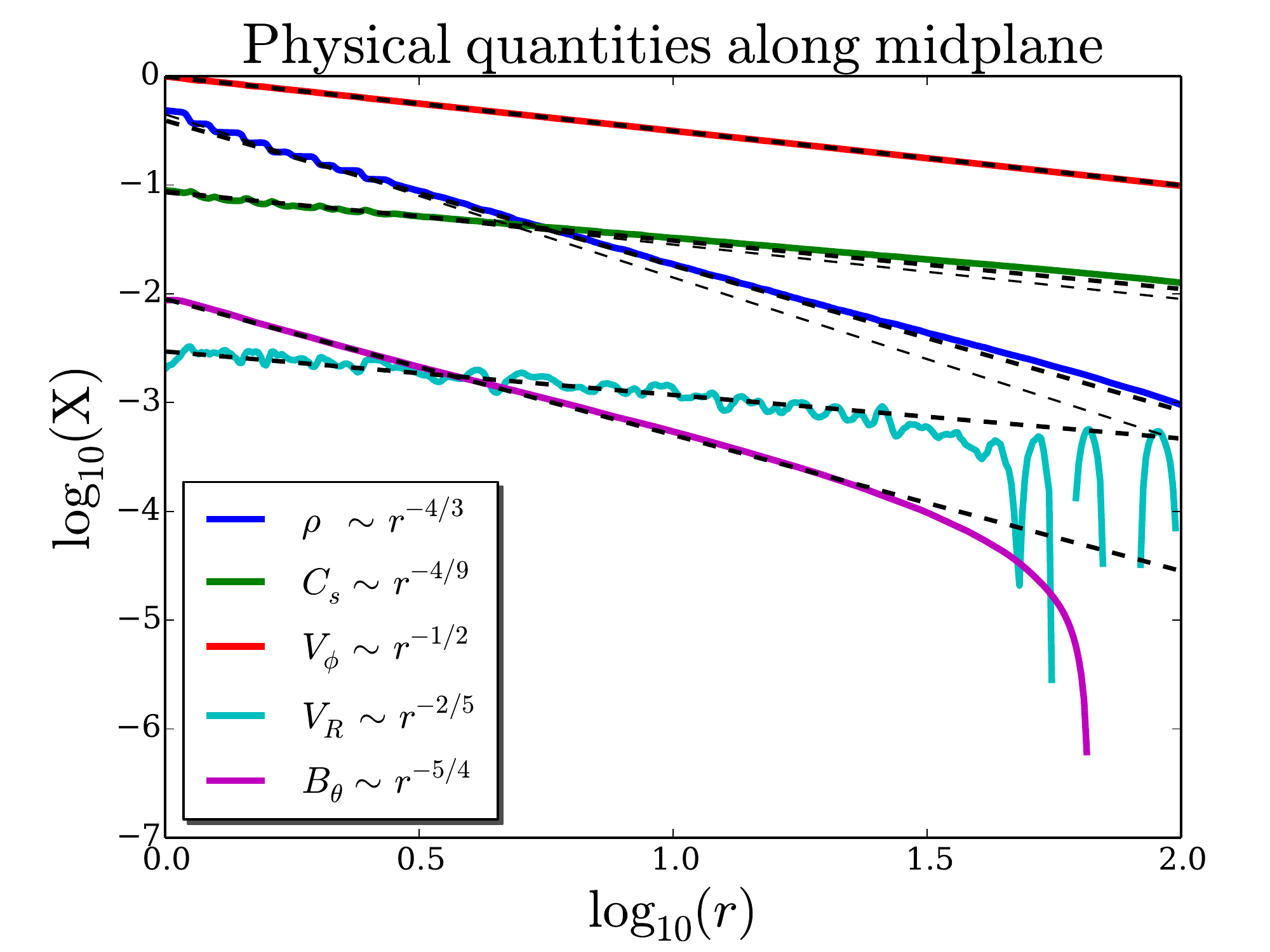}
\caption{Physical quantities along the disk midplane for the reference dynamo simulation at $T=10,000$.
Colored lines correspond to different physical quantities, 
density $\rho$, sound speed $C_{\rm s}$,
rotational velocity $V_{\phi}$, radial velocity $V_{R}$, and the magnetic field component $B_{\theta}$.
The thick dashed lines show the corresponding approximation by a power-law.
The thin dashed lines show the initial power-law distribution, slightly offset from
the actual distribution at $T=10,000$.
}
\label{fig:dyn_profiles}
\end{figure}

Here we discuss the overall disk structure of the reference dynamo simulation.
Figure~\ref{fig:dyn_profiles} presents the radial profiles of a number of MHD variables 
measured at the disk midplane, and the fits to them by power-laws.
The slight deviation between these lines shows how the disk structure changes after a 
long time evolution.
At time $T=10,000$ we find distinct power-laws for the profiles along the disk for radii up to $R \leq 40$. 
This is the radius that marks a steady state area from the rest of the disk,
where the magnetic field is continuously generated.
This is most easy to infer from the profile of the poloidal magnetic field profile, 
that starts deviating from the approximated power-law at $R=40$.

In order to better compare the analytical fits to the radial profiles resulting 
from the dynamo simulation to those without the dynamo (Paper I), we plotted the fits 
with the same power-law indices as for Paper I.
We see that the actual profiles (for example, the density) show only a tiny deviation.

At time $T= 10,000$, we find the following numerical values for the power-law coefficients 
$\beta_X$ for different variables at the midplane $X(r, \theta = \pi/2) \sim r^{\beta_X}$.
The disk rotation remains Keplerian over the whole time evolution, therefore $\beta_{\Vphi} = -1/2$.
The radial profiles for density and thermal pressure slightly change from their initial distributions.
The power-law index of the density changes from  $\beta_\rho = -3/2$ to about $\beta_\rho = -4/3$, while
for the pressure it changes from $\beta_{\rm P}= -5/2$ to about $\beta_{\rm P}= -20/9$.
We find $\beta_{\Vr} = -2/5$ for the accretion velocity, and $\beta_{\Bth} = -5/4$ for the 
poloidal magnetic field.
The accretion velocity remains subsonic everywhere in the disk with an accretion Mach number of 
$\Mr \equiv V_R / \Cs \simeq 0.08$.
As expected, we find strong fluctuations in the area where the dynamo is active and
field magnetic generation is ongoing.

Following \citet{1995A&A...295..807F} and considering the mass accretion rate
$\Macc \sim R^2 \rho \Vr$, it is easy to derive the ejection index { $\xi = 2+ \beta_\rho +\beta_{\Vr}$\footnote{ Steady state and a power-law nature of an accretion rate is implicitly assumed}}, that is a measure of the efficiency of the outflow.
We find $\xi = 0.25$, about the same value as 
for the non-dynamo simulation (see Paper I).

In this section we have demonstrated that the radial profiles for disk dynamics along the midplane
are very similar for the simulations including dynamo action and those without dynamo (in Paper I).
We find several reasons for explanation.
First, in the case of moderately weak magnetic field, the disk dynamics is primarily governed 
by the hydrodynamical quantities, but not so much the magnetic field.
The power-law nature of the Keplerian rotation dominates the dynamics and forces the other 
hydrodynamical profiles into a power-law distributions as well.
Second, the magnetic field strength resulting from the disk evolution is of the same order for both
approaches, $\mu \approx 0.01$, that also can lead to profiles with to the same distribution.

\subsection{Dynamo Versus Non-dynamo Simulations}
Here we discuss the major differences between the simulations with and without the mean-field $\alpha^2\Omega$ dynamo.
As pointed out in the previous sections, the major difference between dynamo and non-dynamo simulations
is the structure of the magnetic field and not so much the accretion disk hydrodynamics.
The dynamo generates the magnetic field that is continuously spreading over the whole disk.
In the early stage of the disk evolution, this makes a substantial difference.
Later, when the inner part of the disk has reached a steady state, this part of the
disk looks very much the same except a few details.
One difference can be found concerning the disk wind close to the disk surface.
In dynamo simulation the sonic surface and the Alfv\'en surface are located 20\%-30\%
further up into the outflow.
However, the magnetic lever arm (radius of the Alfv\'en point) is about the same.
This is the result of the vertical component of the magnetic field being stronger, 
respectively the lower inclination of the magnetic field with respect to the disk surface.
(more inclined toward the disk surface).
The launching area, namely the area where the velocity field changes from being almost 
perpendicular to the magnetic field into a direction parallel to the field (see Paper I),
is now wider, while the disk surface stays at about the same level.
Note that because of the evolving loop-like structure of the magnetic field,
the field inclination with respect to the disk or the sonic surface does change
in time and space - except the inner disk where a steady-state has established.

In order to study the jet properties with respect to the actual disk magnetization, we have performed 
several simulations, varying the $\MUzero$ parameter in the definition of the diffusivity 
(Equation~\ref{eq:strongdiff}). 
This parameter indirectly governs the resulting disk magnetization, as discussed above and also in Paper I.
By running simulations with different $\mu_0$, we were able to probe the resulting actual 
magnetization of the poloidal component in the inner disk over a range $\MUact = 0.01-0.05$.
As shown in Paper I, it is in this range of magnetization where a change in the dominant 
launching mechanism takes place.

In comparison, we find that in our dynamo simulations the disk and jet properties 
do not really differ for different actual magnetization, and we cannot disentangle 
different launching mechanisms in the dynamo simulations.
In contrast, the disk quantities as well as the jet integrals (see Paper I) behave rather
similar in this range of magnetization for both simulations.

In order to disentangle the causes for this similar evolution, we may recall two points.
The main reason why we can distinguish two different mechanisms in the non-dynamo simulation 
is the ability to generate a strong magnetic shear with sufficiently weak poloidal magnetic field.
This is possible because the diffusivity in the standard model depends 
only on the poloidal magnetic field.
Thus, for a weak poloidal field, the diffusivity is also small, that helps to sustain the strong 
magnetic shear.
In contrast, in the current study, the magnetic diffusivity depends on the average {\em total} 
magnetic field in the disk.
Therefore, a strong magnetic shear (a stronger toroidal magnetic field) directly increases the
diffusivity, and, as a consequence, limits the magnetic shear.

This more subtle interrelation between the simulation parameters and physical processes
again emphasizes the impact of the magnetic diffusivity model applied.

\section{Episodic Jet Ejection Triggered by a Time-variable Disk Dynamo}
In this section we present simulation results of a toy model applying a time-dependent 
dynamo action.
Our motivation is the following.
The dynamo is intrinsically a stochastic phenomenon that in real accretion 
disks can be subject to strong fluctuations.
Some accretion disks may exist in which the dynamo action is suppressed,
while in other disks if certain conditions are met the dynamo can start to operate. 
Also, dynamo quenching mechanism can stop an already working dynamo, and thus lead
to a reconfiguration of the disk-jet system.
Here we refer to the well-known solar cycle as an example.
It is believed that a solar dynamo is responsible for the reconfiguration of the magnetic field
of the Sun.
The strong toroidal field component reveals itself as sunspots with a cyclic appearance.
This solar periodicity can be understood as triggered by a constantly operating dynamo (or maybe different 
types of dynamo).
Another interesting feature of the solar activity in this respect is that it exhibits long-term minima
\citep{1976Sci...192.1189E}, during which there were no sunspots observed at all.
It is believed that during these minima the dynamo action is either strongly suppressed or completely 
switched off.

\begin{figure}
\centering
\includegraphics[width=9cm]{\figurepath/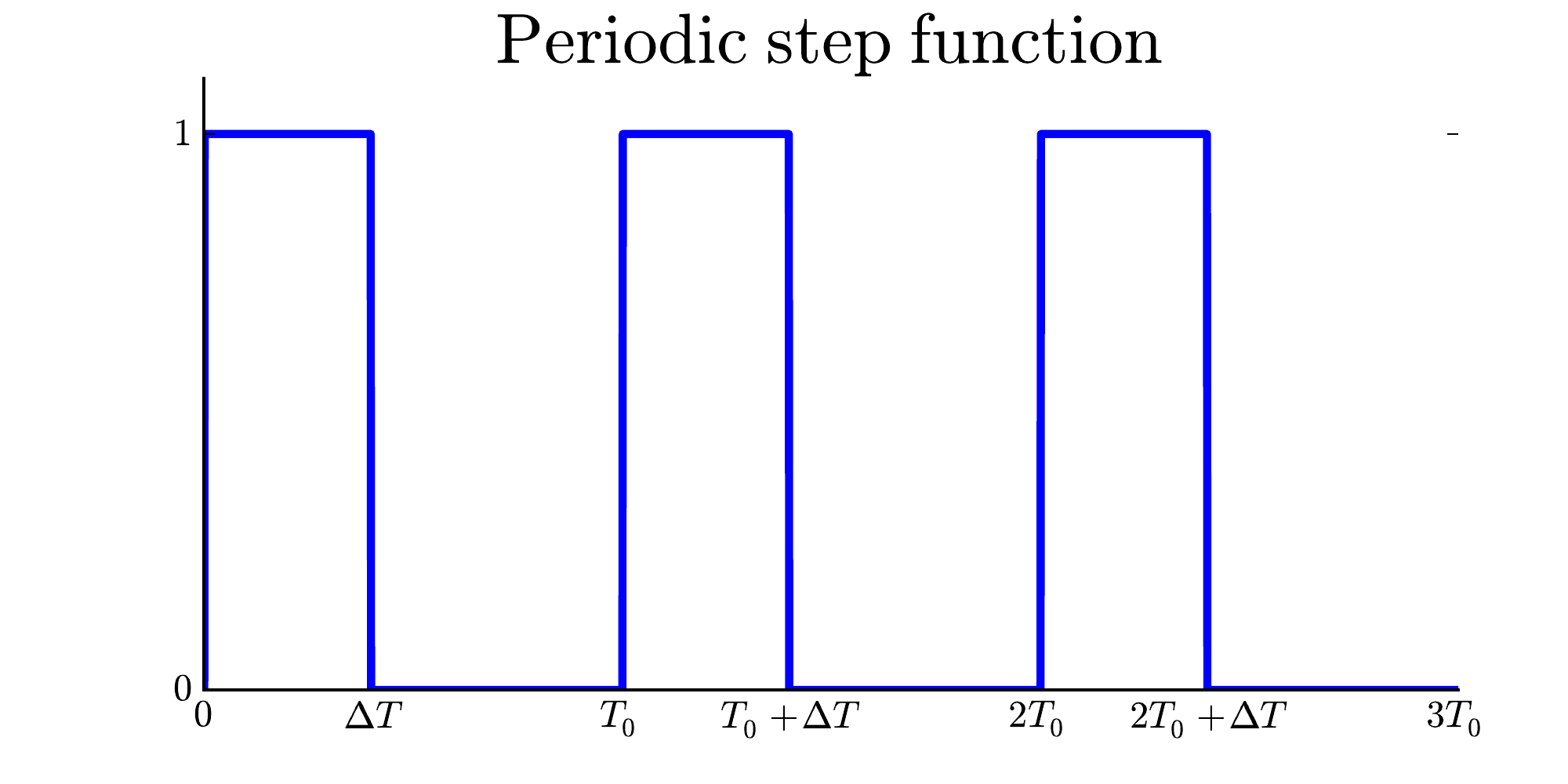}
\caption{Periodic step function applied for the toy model of a time-variable disk dynamo.
Here the dynamo alpha is simply modulated by the periodic step function.
Thus, the dynamo is switched on after periods of $n T_0$ and switched off at $n T_0 + \Delta T$. 
}
\label{fig:steps}
\end{figure}

For our accretion-ejection simulations we follow a preliminary approach and apply a simple toy model to 
explore the impact of such an effect for jet-launching.
We multiply the spacial $\alpha$ profile with a time-dependent function.
Here we apply a periodic step function (Figure~\ref{fig:steps}), by which we continuously switch on 
and off the dynamo in the disk. 
The periodic step function is characterized by its period $T_0$ and time length of a step function $\Delta T$.
In other words, $T_0$ is the period of the dynamo cycle and $\Delta T$ the
activity cycle of the dynamo.
Below we present the simulation with $T_0 = 1000$ and $\Delta T = 400$.

\begin{figure}
\centering
\includegraphics[width=9cm]{\figurepath/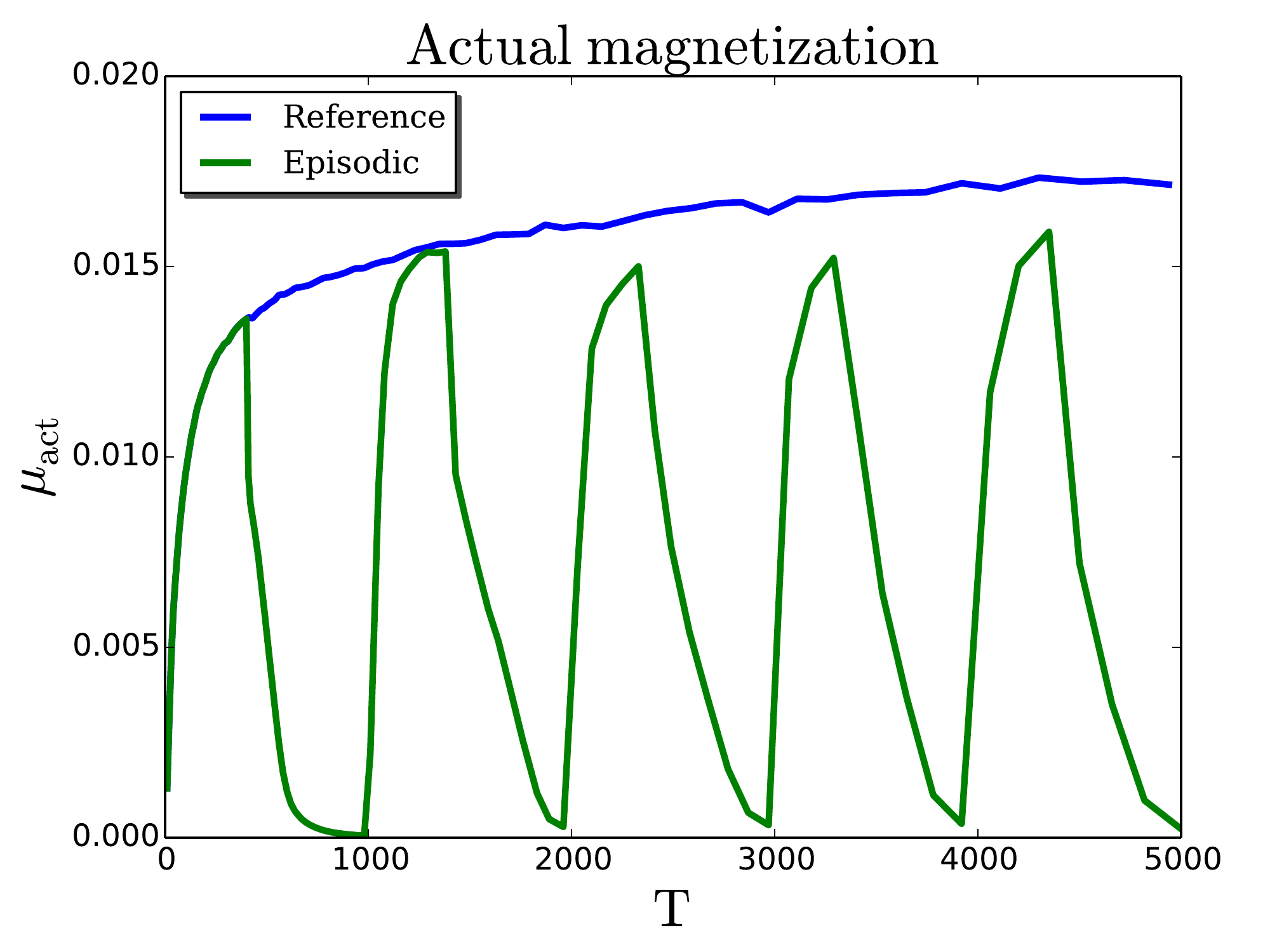}
\caption{The time evolution of the actual disk magnetization $\MUact$ of the reference dynamo simulation 
(blue line) and time-dependent dynamo simulation (green line).}
\label{fig:dyn_tmu}
\end{figure}

\begin{figure}
\centering
\includegraphics[width=9cm]{\figurepath/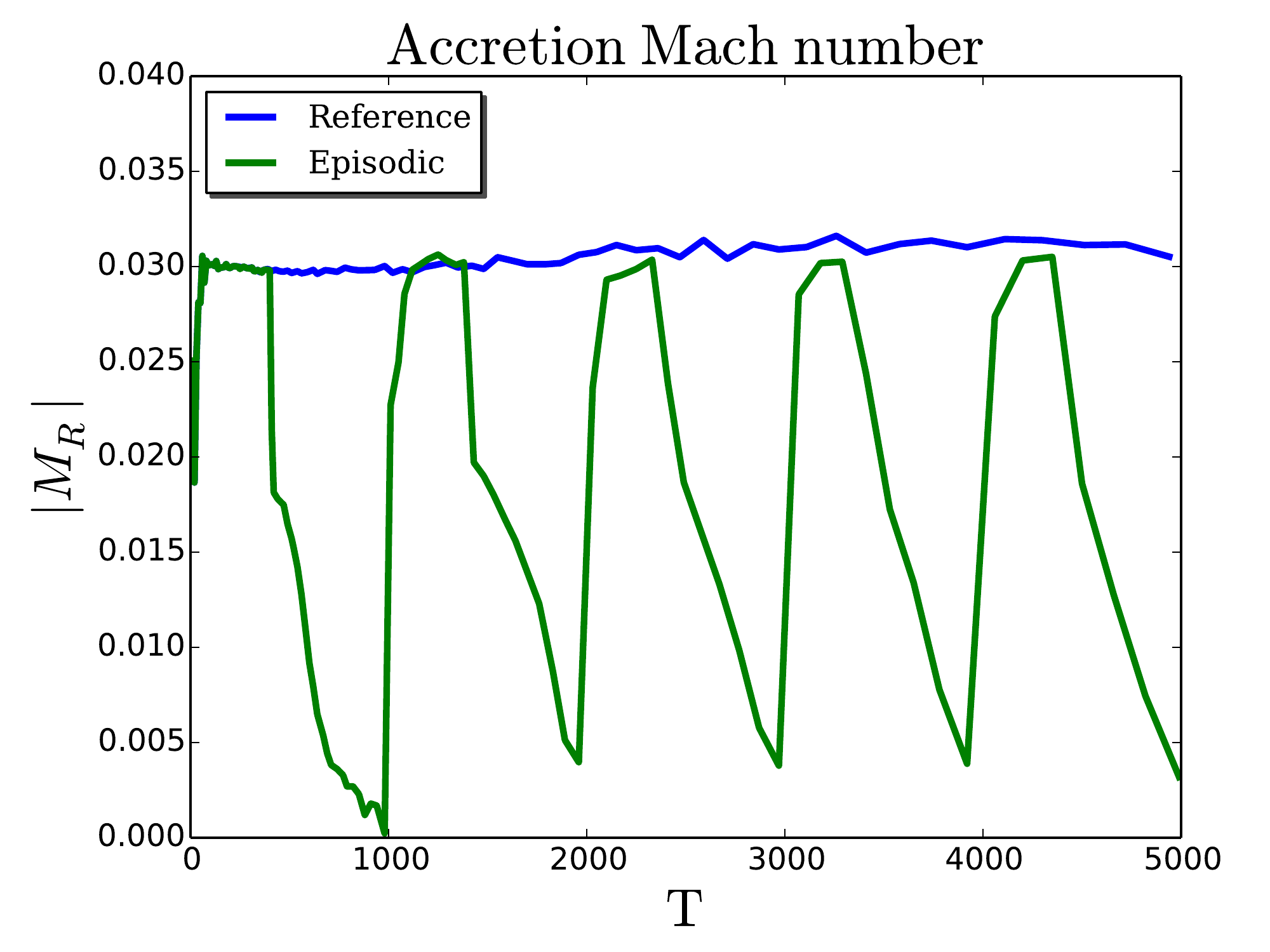}
\caption{The time evolution of the accretion Mach number $\Mr$ of the reference dynamo simulation 
(blue line) and time-dependent dynamo simulation (green line).}
\label{fig:dyn_tmr}
\end{figure}

\begin{figure*}
\centering
\includegraphics[width=18cm]{\figurepath/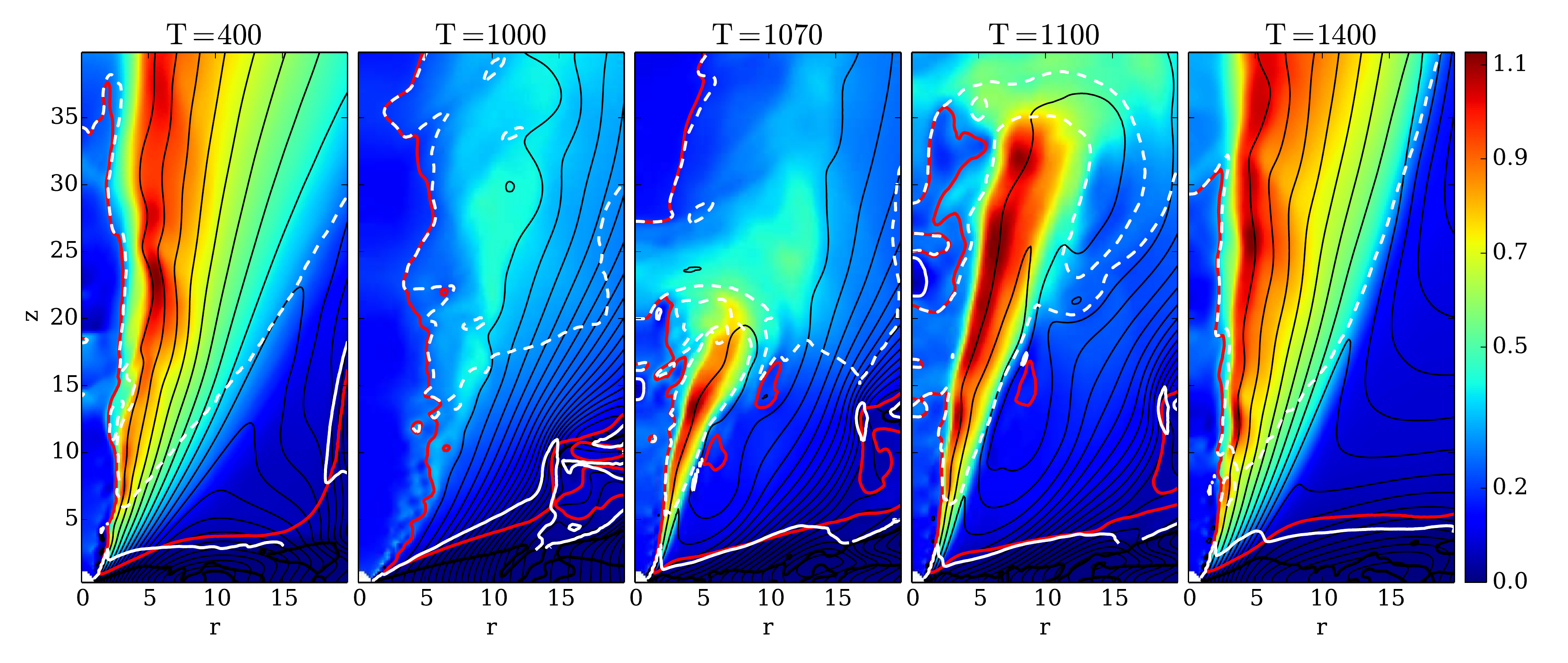}
\caption{Time evolution of the disk-jet structure of the simulation with time-dependent dynamo action.
Only a small cylindrical part of the whole spherical domain is presented.
Shown is the { velocity (colors)};
the poloidal magnetic field lines (thin black lines); 
the sonic (red line), the Alfv\'en (white line), and the fast magnetosonic
(white dashed line) surfaces.}
\label{fig:epi_tevol}
\end{figure*}

Essentially, the modulation of the dynamo-$\alpha $ leads to the variation of the disk magnetization 
(Figure~\ref{fig:dyn_tmu}) with the same periodicity.
The strength of diffusivity is chosen such that without a dynamo working in the disk, the advection 
of the magnetic flux with the accreted material cannot balance the outward diffusion of the
magnetic field. 
This eventually results in a decrease of the disk magnetization.
As previously shown (see Paper I for details), there exists a limit of the strength of the
magnetization below which the disk cannot sustain a jet.
When the disk magnetization decays below the level of $\mu \approx 10^{-3}$ the strong jet disappears.
On the other hand, if the dynamo action is re-established, the disk magnetization
grows again and the outflow is re-launched.

We performed a series of parameter runs, varying both $T_0$ and $\Delta T$.
In principle, different scenarios are possible, depending 
not only on the period of step function, but also on the magnetic diffusivity and 
the dynamo parameter (and various combinations of those).

In order to generate episodic ejection events, thus in order to re-establish a jet-driving magnetic field 
structure, several conditions have to be met by the dynamo process.
First, the dynamo must be strong enough in order to generate the magnetic field sufficiently fast.
Second, in order to suppress the jet ejection during consecutive switch-off periods, $(n T_0 - \Delta T)$,
these periods should be sufficiently long, and/or the magnetic diffusivity should be sufficiently high.
Only when the inner disk magnetization decreases below $\mu \approx 10^{-3}$, the jet-launching can no be sustained and the strong outflow disappears.
The overall evolution of these processes depends both on the periods $T_0$ and $\Delta T$ of the step 
function. 

The interplay between dynamo action, accretion and diffusion may lead to different scenarios for the 
episodic ejection events.
If the switch-off period of the the dynamo is shorter than the timescale for the magnetic field to diffuse 
out, the jet ejection will be constantly sustained, and the jet mass and energy fluxes will be just modulated.
If the disk magnetization decays below the value that is necessary to drive a jet, and if the dynamo is weak 
or works only for a short time, the magnetic field will not be re-established adequately, and, therefore
no new jet will be ejected.

Applying such a toy model we are able to produce {\em episodic jet events}, during which the jet as 
well as the disk variables undergo substantial changes.
The change of the disk magnetization directly affects both accretion and ejection processes.
As the magnetization varies in time, also the other physical quantities vary.
As discussed in Paper I, the accretion Mach number $\Mr = \Vr/\Cs$ is tightly related to the disk
magnetization. 
Figure~\ref{fig:dyn_tmr} clearly shows that variations in the disk magnetization triggered by the
time-dependent dynamo directly affect the disk accretion.

Figure~\ref{fig:epi_tevol} shows the time evolution of a time-dependent dynamo simulation.
As for the case when we discuss the evolution of the reference simulation (Figure~\ref{fig:dyn_tevol}), 
we show only a small cylindrical part of the whole spherical domain.
We kept all parameters the same as in the reference dynamo simulation, just folding
the dynamo term with the periodic step function.
Obviously, compared to the reference dynamo simulation, the overall structure of the disk-jet system 
changes in time - substantially, and not smoothly as for the case of a constant dynamo effect.
The dynamo is working until $T=400$ when it is switched off. 
Therefore before $T=400$ the evolution of the disk and outflow was identical
to previously discussed (see Figure~\ref{fig:dyn_tevol}).
Between $T=400$ and $T=1,000$ the generation of the magnetic field by the dynamo was switched off. 
As a consequence, the magnetic field substantially diffuses and the jet velocity decreases.
Although the disk magnetization is continuously decreasing, a weak outflow is still present.
It is the period when the dynamo is switched off ($T_0 - \Delta T$), that indirectly limits the disk magnetization.
The more time is given to diffuse away the magnetic field, the smaller the resulting disk 
magnetization will be in the period when the dynamo is switched off.

At $T=1,000$, the dynamo is switched on again, and the generation of the magnetic field is
re-established.
Because the dynamo-$\alpha$ is rather high and a substantial magnetic flux has remained from the 
previous cycle, it takes a rather little time to reach again sufficient magnetization for strong 
outflow launching.
Once the substantial magnetization of the disk is reached, $\mu\approx 10^{-3}$, the outflow is re-launched.
 from the inner part of the disk, the outflow re-establishes to the outward direction.
Advection of the magnetic flux, together with the accretion material leads to amplification of the magnetization.
At $T=1,400$ the typical magneto-centrifugal structure of the magnetic field is re-established and
a quasi-steady outflow re-appears, thus closing the activity cycle.
Essentially, these magnetic cycles and subsequent jet ejection are repetitive.
However, they are not necessarily identical, due to the magnetic field structure remaining 
from the previous cycle.

We note that the dynamo mechanism discussed above is able to regenerate the magnetic field in 
the disk completely.
This is indeed different from the case of a simple modulation by the change in the magnetic diffusivity 
parameter $\ass$. 
In Paper I we have shown that the diffusivity parameter $\ass$ is very crucial for determining the 
{\em actual} disk magnetization.
Without a disk dynamo, we were able to modulate the outflow simply by varying $\ass$ parameter, 
however, it was impossible to drastically affect the structure of the magnetic field, as we can
now do by the dynamo.

\subsection{Structure and Evolution of the Episodic Jets}
\begin{figure*}
\centering
\includegraphics[width=15cm]{\figurepath/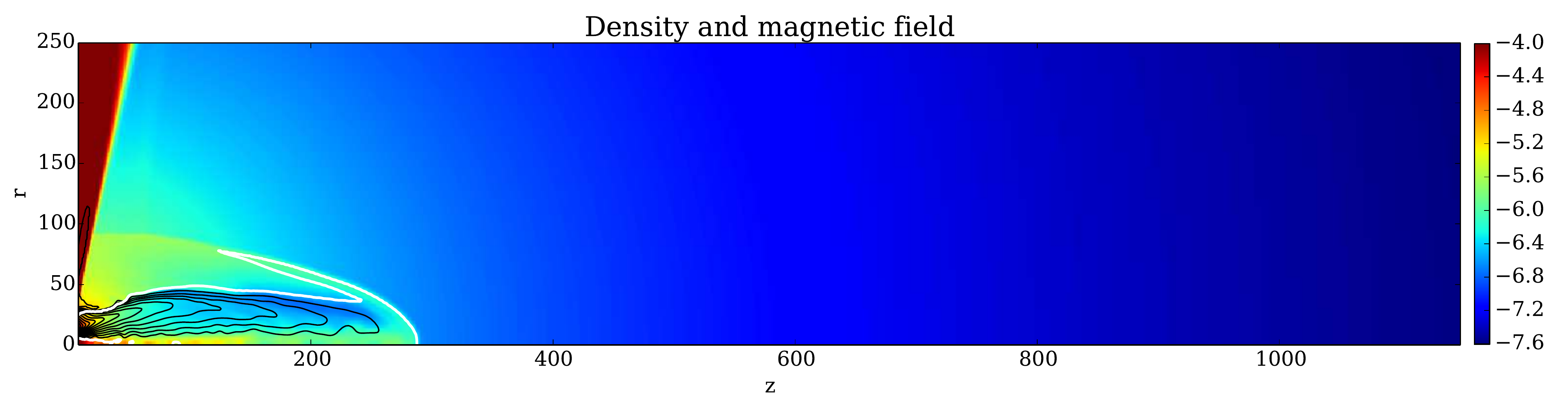}
\includegraphics[width=15cm]{\figurepath/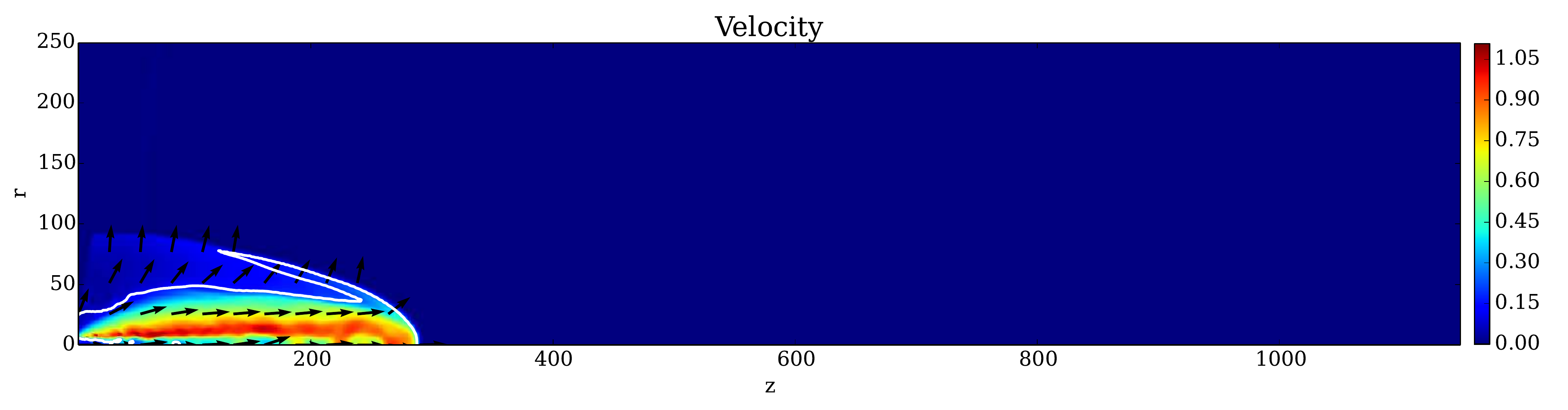}
\includegraphics[width=15cm]{\figurepath/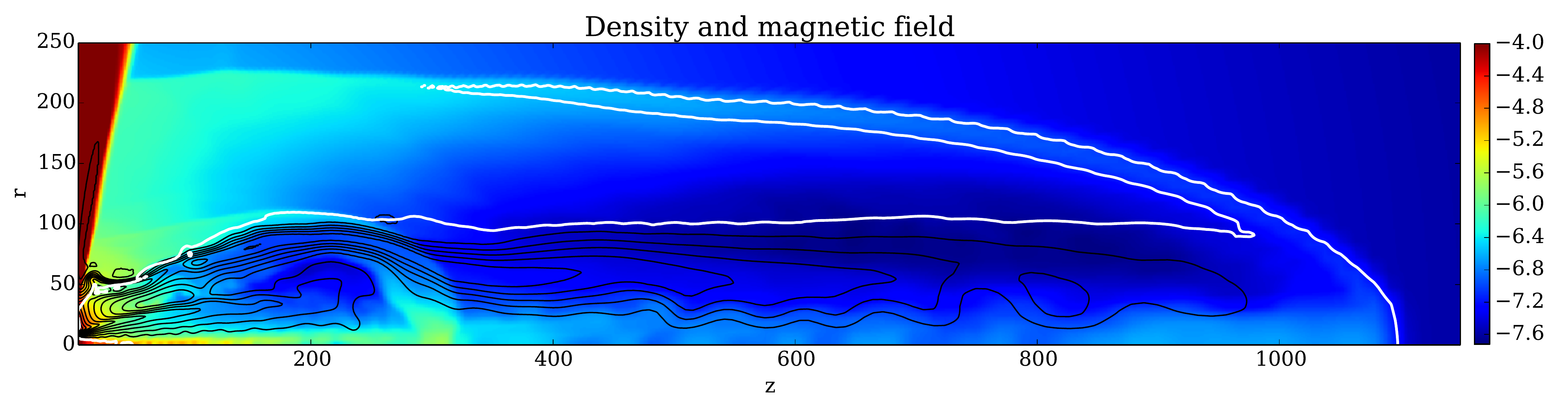}
\includegraphics[width=15cm]{\figurepath/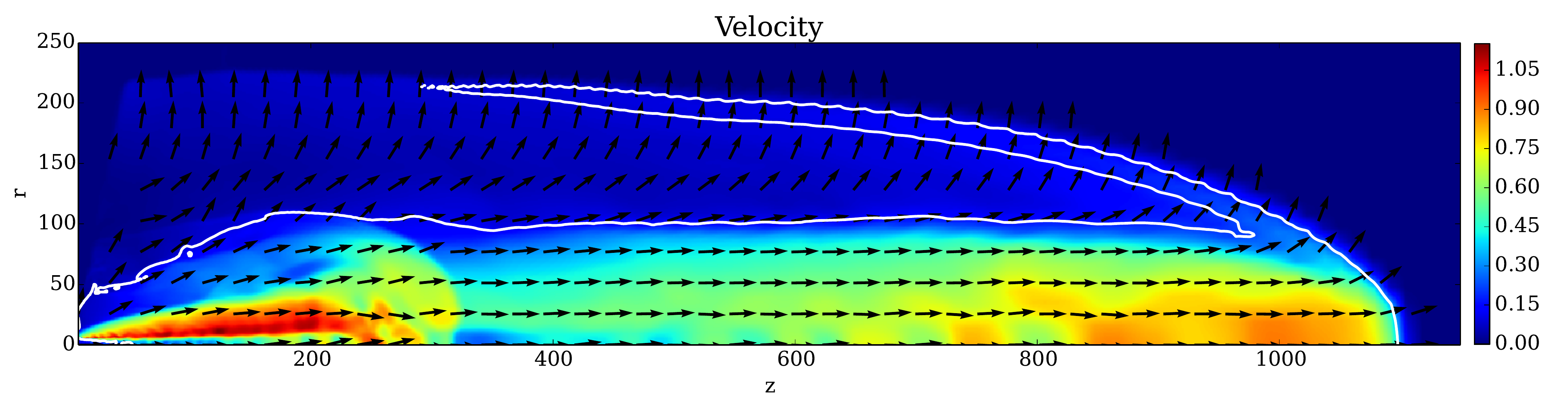}
\includegraphics[width=15cm]{\figurepath/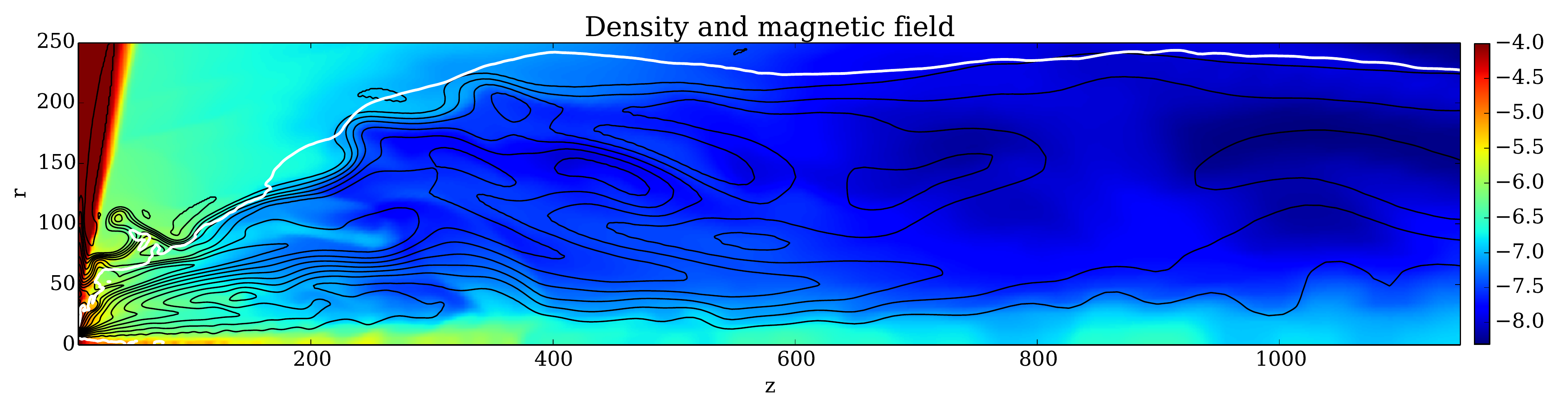}
\includegraphics[width=15cm]{\figurepath/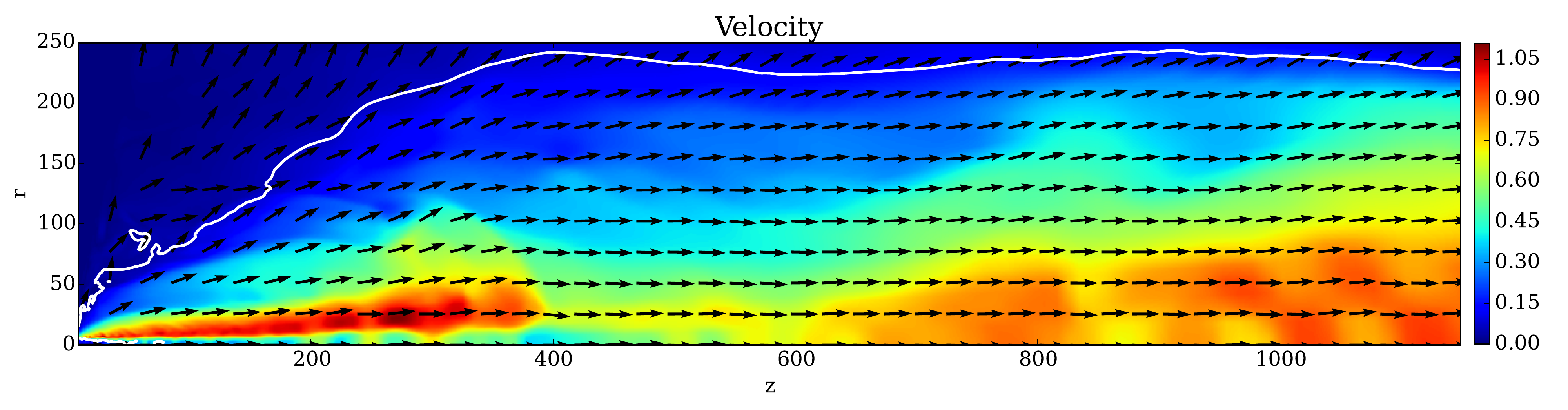}
\caption{Structure of the simulation with time-dependent dynamo action shown at $T=(450, 1450, 3450$), from top to bottom.
Plots are grouped by two.
Shown by color density logarithm (upper plot) and jet speed (lower plot).
Maximum density is set to $10^{-4}$.
Black lines show the magnetic field.
The white line represents the sonic surface.
Arrows show the normalized velocity vectors.
}
\label{fig:knot_struct}
\end{figure*}

%
%
Figure~\ref{fig:knot_struct} presents a time series of snapshots of the high-speed 
outflow propagating close to the axis.
Again, we show only a small cylindrical part of the whole spherical domain, choosing
time and space scales in order to display on the main outflow features.
In order to show typical stages of the episodic ejection generation and propagation we have chosen the 
three dynamical time steps times $T= 450, 1,450, 3,450$ of our simulation lasting in fact
10,000 dynamical time steps.

First, we see how the outflow is generated initially (thus from the first cycle of dynamo
activity) and then propagates throughout the hydrostatic corona.
After switching off the dynamo at $T=400$, the jet weakens until it almost completely disappears.
The dynamo becomes active again at $T=1,000$.
At $T=1,450$ that first ejection has already reached $R \approx 1,100$, just when another "jet" has 
been launched.
At even later time, $T=3,450$, this second knot has established an outflow, and a third knot is
launched.

The timescales we have chosen are such that we can follow multiple ejection events on our grid. 
The time $T=1,000$ would correspond to about $T/2\pi$ inner disk rotations, thus
about a year if we apply the inner disk radius $R_0 = 0.1\,$AU.
For comparison, jet observations of young stars suggest the timescales between the knots
$\tau_{\rm knot} \simeq \Delta L / v_{\rm jet} \sim$ of about 10 years.
If the knot generation mechanism is indeed triggered by a disk dynamo, the timescale
for the field reversals must be longer.


In the two middle panels of Figure~\ref{fig:knot_struct} we clearly see two fast rapidly
moving gas ejections.
These parcels of ejected material are separated from each other by the typical period of 
the dynamo $T_0$, corresponding to about $1000 \ri$ distance between them.
The ejected material rams into the gas which is left from the previous parcel and which
moves with lower velocity.
Shocks are generated that can be clearly seen in the density map\footnote{The shock structure
is also visible in the pressure distribution and also in the jumps in the velocity profile along 
the jet (not shown as figure).}.
We may interpret the repeated ejections as jets knots, however, a more detailed (future) 
investigation would be necessary to confirm this picture.

We may clearly identify the signatures of the inflow from the inner coronal boundary into
the domain along the outflow axis for low $z<200$.
As discussed above, this axial inflow is essential to provide the gas pressure that balances the 
collimation forces of the outflow in the vicinity of the disk.
In our simulations it is injected artificially by the boundary condition,
however, an astrophysical interpretation could be that of a wind driven
by the central object.

\subsection{Self-induced Magnetic Field Regeneration}
\begin{figure}
\centering
\includegraphics[width=9cm]{\figurepath/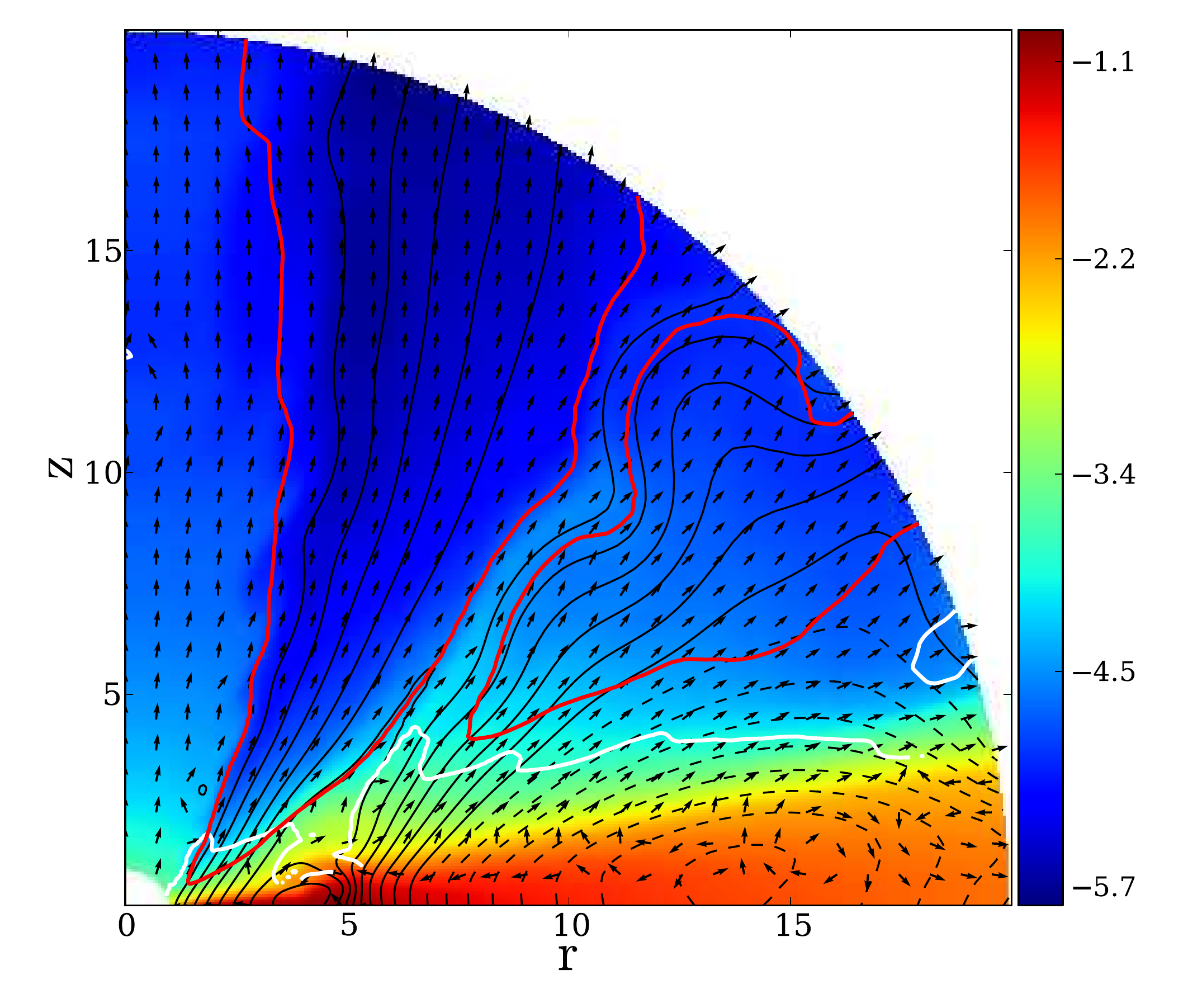}
\caption{Example simulation resulting in two opposite magnetic loops 
generated by dynamo.
Shown is the mass density (colors, logarithmic scale), 
the poloidal magnetic field lines (black lines, dashed lines show opposite magnetic flux),
the sonic surface (red line), and 
the Alfv\'en surface (white line) 
at time $T=1000$. 
Arrows show normalized velocity vectors.
}
\label{fig:dyn_snapshot}
\end{figure}
As discussed in the beginning of this section, the dynamo action is a stochastic, highly non-linear
process.
We can expect that under certain conditions the non-linear evolution of the dynamo is more
pronounced than under other conditions.
In this section we show preliminary results of simulations evolving in a more stochastic way 
and, by that, may be considered as a more natural {"}switch{"} for the dynamo mechanism.
No artificial switch on/off has been applied.
These simulations consider a self-induced regeneration of the magnetic field without applying
any additional constraints such as a periodic step function in the time-dependent dynamo profile.

In our simulations we have observed very similar self-induced regeneration processes 
of the magnetic field under different conditions.
Thus, there seem to be several ways how a self-induced switch of the dynamo regeneration
can take place.
Some of them require the presence of a quenching mechanism.
Without that, the magnetic flux will be contineously generated in the inner disk and will eventually fill 
the entire disk with magnetic field of one dominant polarity.

One possibility to establish a self-induced switch for the magnetic field,
is to initiate the simulation with the disk filled by the toroidal magnetic field of different polarity.
Then the poloidal magnetic field that is generated by the dynamo will have different polarities as well.
Constant field amplification will lead to the accretion/advection of this magnetic structure.
As these structures move toward the center, the generation of the magnetic field in the innermost
structure will be quenched, while the magnetic flux in the outer disk will continue to grow.
As a consequence these structures merge (by diffusion) and decay, and quiescent period of 
outflow launching follows.

Another way is to link the dynamo term to the toroidal magnetic field.
As discussed above, the poloidal magnetic field in the outer disk has an opposite polarity with
respect to the inner disk magnetic field.
Thus, in this case an additional feedback channel is provided that - under certain circumstances -
can lead to a more fluctuating evolution of the disk-outflow system.
The last example that surprisingly showed such self-induced regeneration of the magnetic field is
our reference simulation, but with lower dynamo-term $\ALdisk=-0.03$.
Figure~\ref{fig:dyn_snapshot} shows the magnetic field structure in the disk of one of such simulation.
We have observed that sometimes the dynamo generates several magnetic field loops in the disk.
While the magnetic field in the inner disk is able to quench the dynamo, 
in the outer disk the magnetic field is continuously amplified.
The magnetic flux generated in the outer disk is of opposite polarity.
If advected inward, it will eventually reconnect with the magnetic flux in the inner disk. 
During this cancellation (reconnection) process, the disk magnetization in the jet 
launching region will decrease below a critical level, jet-launching will decay, and the 
outflow will disappear.
At later stages, when the magnetic field remaining from the reconnection process becomes 
sufficiently amplified by the dynamo, the outflow will be launched again.

The details of the process of magnetic field regeneration in fact depends on many model parameters, 
in particular the magnetic diffusivity model.
Although this might be an interesting mechanism triggering episodic events, we do not present 
details here, since we were not be able yet to get it work robustly, and run the simulation longer than
for just a few regenerations.

\section{Conclusions}
We have presented results of MHD simulations investigating
the generation of the magnetic field by the accretion disk dynamo in the context of 
jet and outflows launching.
The time evolution of the disk structure is self-consistently taken into account.
The simulations are performed in axisymmetry treating all three field components.
We apply the MHD code PLUTO-4.0, 
that we have modified for the mean-field $\alpha^2\Omega$ dynamo 
problem in the induction equation.

In the present work we explored the generation of a large scale, global magnetic field.
Our simulations were initiated by the purely radial magnetic field with magnetization $\MUinit = 10^{-5}$.
We showed in detail how the magnetic field is being generated
and through which consecutive stages it evolves, acquiring
in the end the ability to launch jet and outflows.
In this respect our simulations can be seen as a continuation of early work by 
\citealt{2003A&A...398..825V} and \citealt{2004A&A...420...17V}.
In our paper we are concentrating more on the jet and outflow generation and propagation.

One advantage of our simulations is that our model keeps the disk magnetization 
at a rather low level.
Therefore, the magnetic field does not substantially affect the disk hydrodynamics,
and allows to evolve our simulations for very long time.
Each simulation has been evolved at least up to $T=10,000$ on a spherical domain with $R=[1, 1500]$.

In the following, we summarize our main results.

(1) In our simulations treating a mean-field disk dynamo, we may distinguish two main 
features in the magnetic field structures.
The magnetic field of the inner disk that is similar to the commonly found open field structure,
favoring a magneto-centrifugal launching of the outflow.
The poloidal magnetic field of the outer part of the disk is highly inclined, and predominantly radial. 
Differential rotation induces a very strong toroidal component from it. 
Such a structure is similar to what is known as tower jet or Poynting jet in literature.
In this part of the disk, it is toroidal magnetic field that plays a key role in outflow launching.
First, below the disk surface ($\Vr = 0$) the matter is lifted by the buoyant force of the magnetic field, thus, by the gradient of the thermal pressure.
Starting from the disk surface ($\Vr = 0$), the matter is further accelerated by the pressure gradient of the toroidal magnetic field.
The outflows from the outer part of the disk are typically slower, denser,
and less collimated, thus corresponding to a higher mass loading.

(2) In principle, the dynamo can fill the entire accretion disk with magnetic flux.
Thus, if the dynamo action is not quenched, magnetic flux is continuously generated, diffuses
outward along the disk until it fills the entire disk.
This loop-like structure of the magnetic field that is typical for a dynamo, propagates
further outward.

(3) As soon as the disk magnetization reaches a critical limit, $\mu > 10^{-3}$, disk winds are
launched and can be accelerated to super-magnetosonic speed.
This result is complementary to our earlier simulations that do not consider dynamo \citep{paper-1}, and 
where the critical magnetization was obtained just from advection of magnetic flux by accretion.
Thus, again we can confirm the long-standing belief that the disk magnetization plays 
the key role in the outflow launching.
In the inner disk, the rate of generation of the magnetic field by the dynamo is higher,
leading to a strong gradient of the disk magnetization.

(4) We have further invented a toy model triggering a time-dependent efficiency of the
mean-field dynamo.
In that model approach, we periodically switch on and off the dynamo.
This strongly affects the magnetic field structure.
The decay of magnetic flux by diffusion can be completely balanced by the dynamo that regenerates
the magnetic field.
As a consequence, the duty cycles of the dynamo action can lead to episodic jet ejection, just
depending on the disk magnetization obtained during dynamo activity.
When the dynamo is suppressed and the disk magnetization falls below a critical value, 
$\mu \approx 10^{-3}$, the generation of outflows as well as the accretion
is substantially inhibited.
We had chosen the timescale of the dynamo cycle and the corresponding timescale for the episodic ejections
somehow shorter compared to the observed values - just because we wanted to follow several
events in the same simulation box.
However, the main - and general - result is that we can steer episodic ejection and {\em large-scale jet knots}
by the {\em disk intrinsic dynamo} that is time-dependent and regenerates the jet-launching magnetic field.

(5) Concerning the disk hydrodynamics, we find that the accretion velocity follows the same power-law
$\beta_{\Vr} \approx -2/5$ for the simulations with and without dynamo.
This interesting also, because we have applied slightly different diffusivity model leading to different 
magnetic field structure.
Nevertheless, the accretion profiles are approximately the same. 
%
As a consequence, we also find approximately the same ejection index $\xi \approx 0.25$.

(6) Although the dynamo and non-dynamo simulations are significantly different, 
launching mechanism of the fast jet is primarily the same.
Thus, from a pure observational point of view, one would not yet be able to distinguish
whether the jets are launched from a dynamo-generated magnetic field or from
a magnetic field advected from the interstellar medium.
%

In summary, we have shown the accretion-ejection evolution considering a 
magnetic field self-generated by a mean-field disk dynamo.
Repetitive ejection could be obtained by a time-dependent dynamo-$\alpha$.
A future step could be to consider the dynamo-action for the strong-field case. 
That might be realized by implementing both a MRI dynamo and a Parker dynamo 
by means of different dynamo-$\alpha$.
Another step would be to have a more direct link between the actual 
magnetic field and the dynamo.

\acknowledgements
We thank Andrea Mignone and the PLUTO team for the possibility to use their code.
The simulations were performed on the THEO cluster of Max Planck Institute for Astronomy.
This work was partly financed by the SFB 881 of the German science foundation DFG.

\appendix

\section{Resolution Study}


We shortly discuss the results of our resolution study.
We have performed simulations with a grid resolution of 0.75, and 1.5 times our standard resolution
of $N_\theta=128$ cells per quadrant, 
corresponding to $N_\theta=96$ and $N_\theta=192$ cells per quadrant, or approximately 12 or 
24 cells per disk height $2\epsilon$
compared to 16 cells per disk height in our reference run\footnote{
Note that once the resolution in $\theta$-direction and the radial extent of the disk is chosen,
the resolution in $R$-direction is uniquely determined (see Section~\ref{sec:model}, or Paper I)}.

\begin{figure}
\centering
\includegraphics[width=5.9cm]{\figurepath/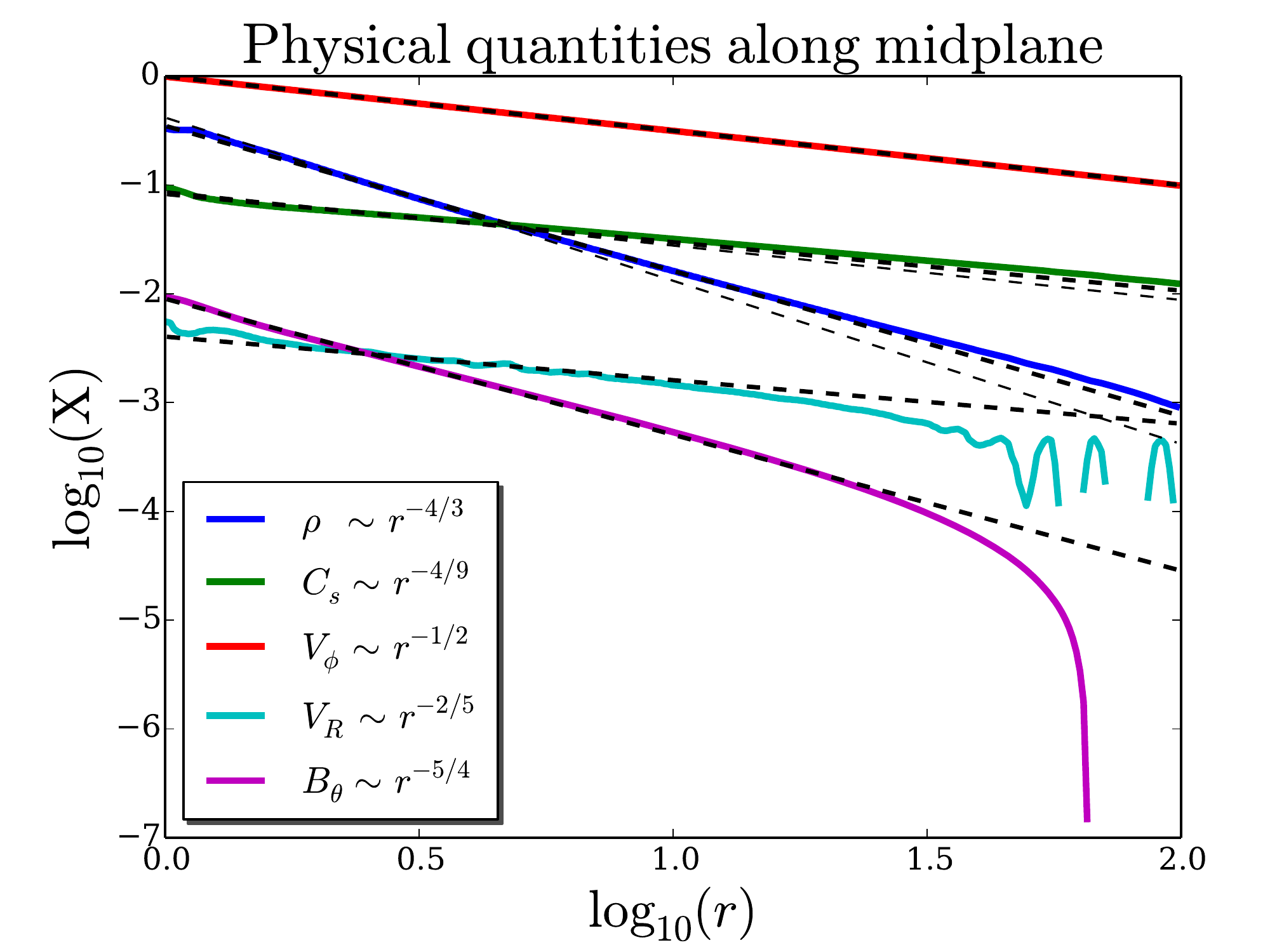}
\includegraphics[width=5.9cm]{\figurepath/res_def_profiles-eps-converted-to.pdf}
\includegraphics[width=5.9cm]{\figurepath/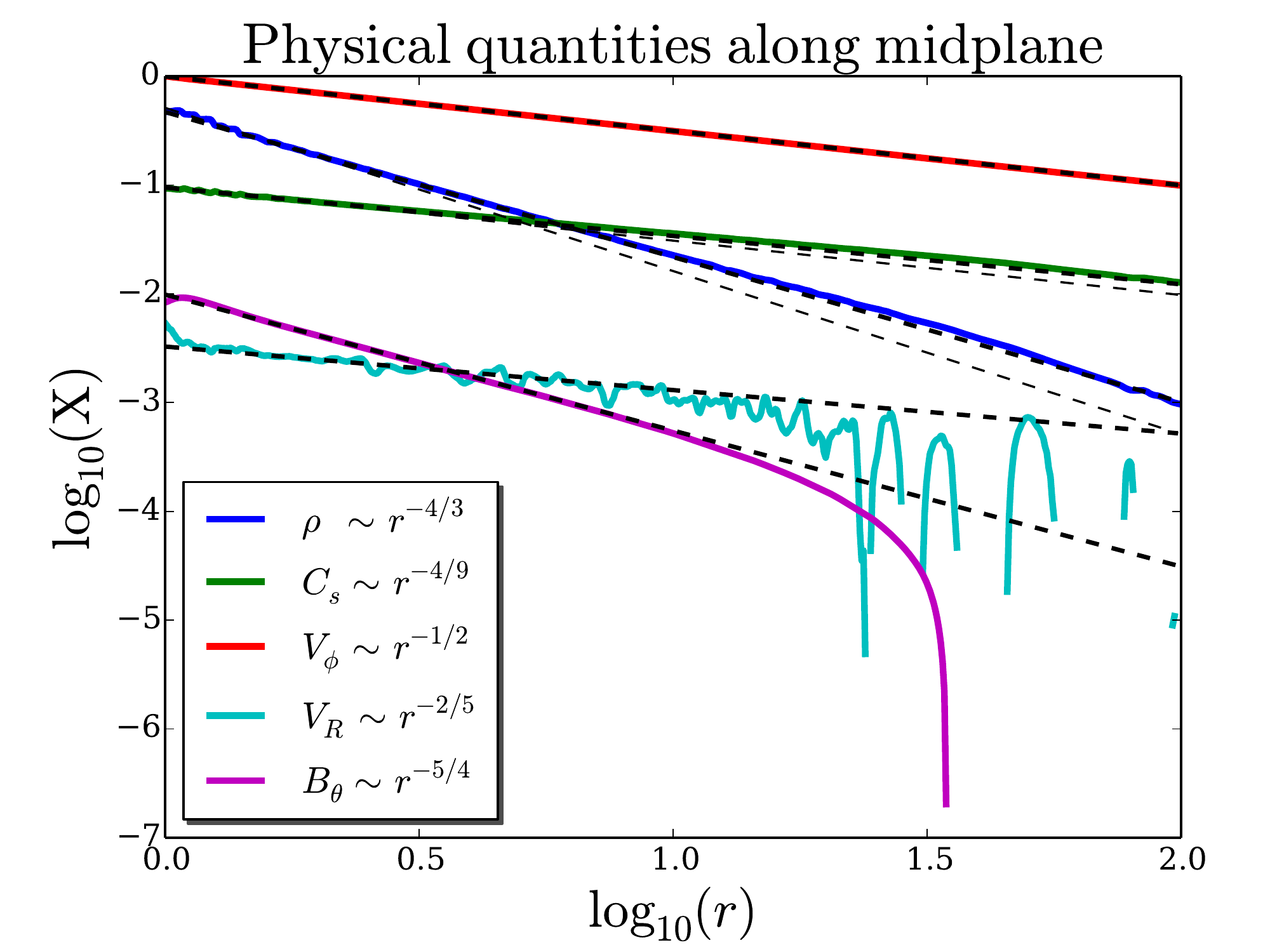}
\caption{Resolution study. 
Physical quantities along the midplane for the simulations
with different resolution at $T=10,000$. 
From top to bottom the resolution is  
(12, 16, 24) cells per disk height ($2\epsilon$).
Colors show
different variable profile, thick dashed lines correspond 
to certain power-law, the mismatched thin dashed lines 
correspond to initial distributions of variables. 
}
\label{fig:res_all}
\end{figure}

Figure~\ref{fig:res_all} shows the dynamical profiles the three simulations of our resolution study.
The radial profiles are plotted along the midplane for various dynamical variables at time $T=10,000$.
As discussed in Section~\ref{sub:profiles} these profiles can be nicely fitted by power-laws.
The same power-law index also provides the same ejection to accretion index.
Therefore, we conclude that our results are not resolution dependent.

However, several differences between these curves can be noticed.
The inner part of the disk for the simulation with lower resolution indicates a substantial 
deviation from the corresponding power-law.
We also find that for a lower resolution the accretion speed increases, and, as a consequence,
the overall density in the disk decreases.
This highlights the effect of the numerical viscosity that enhances the angular momentum
transfer in the disk.
On the other hand, the magnetic field is diffused out faster and to a larger distance. 
This indicates a higher numerical resistivity for the case of lower resolution.

\bibliographystyle{apj}

\end{document}